\documentclass[12pt,english]{article}
\usepackage{
amsmath,
amssymb,
amsthm,
braket,
booktabs,
caption,
cite,
float, 
graphbox,
graphicx,
hyperref,
dsfont,
pgfplots,
stackengine,
tabu,
xcolor,
xparse,
color
}

\usepackage{centernot}
\usepackage{mathtools}
\usepackage{stmaryrd}
\usepackage{bbm}
\pgfplotsset{compat=1.14}
\hypersetup{colorlinks=true,
	linkcolor={red!30!black},
    	citecolor={blue!50!black},
    	urlcolor={blue!30!black}
   	 }
  	 
\newcommand{\eqb}{\begin{equation}}
\newcommand{\eqe}{\end{equation}}
\newcommand{\dmb}{\begin{displaymath}}
\newcommand{\dme}{\end{displaymath}}

\newcommand{\eab}{\begin{eqnarray}}
\newcommand{\eae}{\end{eqnarray}}

\newcommand{\e}{\mbox{e}}
\newcommand{\be}{\begin{equation}}
\newcommand{\ee}{\end{equation}}

\newcommand{\SG}[1]{\ensuremath{\mathrm{SG}(#1)}}
\newcommand{\SU}[1]{\ensuremath{\mathrm{SU}(#1)}}
\newcommand*{\rep}[2][]{\ensuremath{{\boldsymbol{#2}#1}}}
\newcommand{\I}{\ensuremath{\mathrm{i}}}
\newcommand{\Nf}{\ensuremath{F}}

\newcommand{\Nr}{\ensuremath{R}}

\NewDocumentCommand\mat{m}{\left(\begin{matrix}#1\end{matrix}\right)}
\NewDocumentCommand\lagr{}{\mathcal{L}}
\NewDocumentCommand\phir{m}{\phi_{\rep{r}_{#1}}}
\NewDocumentCommand\rhor{m}{\rho_{\rep{r}_{#1}}}
\NewDocumentCommand\ur{m}{U_{\rep{r}_{#1}}}
 \NewDocumentCommand\acurlr{m}{\mathcal{A}_{\rep{r}_{#1}}}
\newcommand{\Z}[1]{\ensuremath{\mathbbm{Z}_{#1}}} 

\definecolor{darkgreen}{HTML}{109930}

\title{\bf Telling compositeness at a distance \\ with outer automorphisms and CP
}
\author{Ingolf Bischer${}^1$, Christian D\"oring${}^{1,2}$, Andreas Trautner${}^1$}

\date{{\it 
${}^1$Max-Planck-Institut f\"ur Kernphysik, \\ Saupfercheckweg 1, 69117 Heidelberg, Germany}\\[0.2cm]
{\it 
${}^2$Service de Physique Th\'eorique, \\ Universit\'e Libre de Bruxelles, C.P. 225, B-1050 Brussels}\\[0.5cm]
\href{mailto:christian.doring@ulb.be}{christian.doring@ulb.be},
\href{mailto:trautner@mpi-hd.mpg.de}{trautner@mpi-hd.mpg.de}
}

\begin{document}
\setlength{\topmargin}{0.0 true in}
{\hfill ULB-TH/22-16\vspace{1cm}}
{\let\newpage\relax\maketitle}
\thispagestyle{empty}

\begin{abstract}
\noindent
We investigate charge-parity (CP) and non-CP outer automorphism of groups and the transformation behavior of group representations under them.
We identify situations where composite and elementary states that transform in exactly the same representation of the group, transform differently under
outer automorphisms. This can be instrumental in discriminating composite from elementary states solely by their quantum numbers with respect to the outer automorphism,
i.e.\ without the need for explicit short distance scattering experiments.
We discuss under what conditions such a distinction is unequivocally possible. 
We cleanly separate the case of symmetry constrained (representation) spaces from the case of 
multiple copies of identical representations in flavor space, and identify conditions under which non-trivial transformation 
in flavor space can be enforced for composite states.
Next to composite product states, we also discuss composite states in non-product representations. 
Comprehensive examples are given based on the finite groups $\Sigma(72)$ and $D_8$.
The discussion also applies to $\SU{N}$ and we scrutinize recent claims in the literature that $\SU{2N}$ outer automorphism with antisymmetric matrices correspond 
to distinct outer automorphisms. We show that outer automorphism transformations with antisymmetric matrices are related by an inner automorphism to the standard $\Z{2}$ outer automorphism of $\SU{N}$.
As a direct implication, no non-trivial transformation behavior can arise for composite product states under the outer automorphism of $\SU{N}$.
\end{abstract}
\newpage
\setlength{\topmargin}{0.4 true in}

\section{Introduction}\label{sec:Intro}
\widowpenalty10000\clubpenalty10000
Detecting the composite nature of previously believed-to-be elementary constituents has a long and successful history in physics,
ranging from macroscopic, over atomic and nuclear physics scales down to the Standard Model (SM) of elementary particles.
Crucial in this endeavor was not only the advent of inelastic scattering experiments, but first and foremost, 
the realization that \textit{composite} states can obey a transformation behavior under the fundamental symmetries of nature 
that often allows to distinguish them from their \textit{elementary} constituents. 

Consider, for example, fermion anti-fermion bound states such as positronium or quark anti-quark mesons. The eigenvalues 
of such a composite state with respect to the transformation under space-time parity P or charge conjugation C are uniquely determined as 
\begin{equation}\label{eq:PCQNs}
 \eta_\mathrm{P} = (-1)^{L+1}\;, \qquad\text{and}\qquad  \eta_\mathrm{C} = (-1)^{L+S}\;,
\end{equation}
where $L$ is the orbital angular momentum of the state and $S$ its total spin. 
This bound state behavior, together with the introduction of (internal) flavor quantum numbers such as isospin, strangeness, charm etc.,
was instrumental~\cite{Gell-Mann:1957khb,Schwinger:1957em} in deciphering the composite nature of mesons and 
baryons~\cite{Gell-Mann:1961,Neeman:1961jhl,Gell-Mann:1962yej}, ultimately giving rise to the 
successful quark model of hadrons~\cite{Gell-Mann:1964ewy,Zweig:1964jf}
some time before the composite nature of hadrons had been confirmed by deep inelastic scattering~\cite{Breidenbach:1969kd}.

The compositeness of an observed state can be established by determining its quantum numbers. 
Since this may take place at a distance, for example by carefully examining decay products, no direct test of the spatially 
extended nature of a state or its constituents, say in a direct scattering experiment, may be necessary in order to establish its compositeness.
Crucially, this depends on the assumption of which states may or may not appear as elementary states in the spectrum of a theory.
For example, assuming the absence of an elementary pseudo-vector (often also called axial-vector) state in Nature,
it is clear that pseudo-vector mesons with quantum numbers $J^{PC}=1^{+-}$ (e.g.\ $h_1(1170)$, $b_1(1235)$, etc.~see~\cite{ParticleDataGroup:2022pth}) \textit{must} 
be composite states. The same can be said for the pseudo-scalar mesons with $J^{PC}=1^{-+}$ (e.g.\ $\pi^0$, $K^0$, etc.). However, 
in the latter case it is more intricate to proclaim compositeness solely based on external quantum numbers, simply because elementary pseudo-scalar fields might exist.
For example, this can be in the form of (pseudo-)Goldstone bosons such as axions or axion-like particles (ALPS), or as heavy pseudo-scalar states of multi-Higgs extensions of the SM.
The necessity of an assumption about the nature of allowed elementary states will also plague our determination of compositeness below. 
This arises already in the most well-known examples and will not hinder our following discussion.

Note that it makes sense to consider P and C quantum numbers for positronium and multi-quark composite states, 
despite the fact that both transformations are explicitly and maximally broken in the SM.
This is because the composite states here are non-relativistic bound states comprised of Dirac fermions
for which P and C transformations are well-defined. 
Furthermore, both transformations are preserved by the vector-like acting gauge interactions of QED and QCD which explains their importance in the classification of states.

For more general situations, e.g.\ relativistic (or massless) states and transition mediated by the heavy electro-weak gauge bosons only the combined transformation of CP 
is well-defined (albeit also broken, but only explicitly not maximally). 
Despite what the naming suggests, actually CP is more fundamental than either of the transformations C and P. 
This can be understood from the fact that all of these transformations are, group theoretically speaking, outer automorphisms
of the gauge~\cite{Grimus:1995zi}, space-time~\cite{Buchbinder:2000cq} and global (e.g.\ flavor) symmetries~\cite{Holthausen:2012dk}
of a theory~(see e.g.~\cite{Trautner:2017vlz} for a concise introduction, as well as~\cite{Fallbacher:2015upf,Trautner:2016ezn} and references therein for the deep dive). 
CP plays an outstanding role because it corresponds to a class-inverting automorphism~\cite{Chen:2014tpa}, which corresponds to a complex conjugation transformation for groups with complex representations. 
While P and/or C can be maximally broken by omitting certain representations (as is the case in the SM) the CP transformation is 
always well-defined for a gauge theory with real valued action. This explains why we put a focus on CP transformations of composite states in the following,
even though our discussion and examples apply to the case of general outer automorphisms. 

As soon as there are multiple fields involved with identical quantum numbers under the preserved symmetries, i.e.\ as soon as there are multiple ``flavors'' of fields,
outer automorphisms such as CP can be generalized to obey non-trivial transformations in flavor space~\cite{Lee:1966ik}.\footnote{%
One should strictly distinguish here between ``general\textit{ized}'' and ``general'' CP transformations. 
The possibility to ``generalize'' a CP transformation arises in situations with a ``free'', i.e.\ symmetry unconstrained, flavor space 
like in the SM. By contrast, ``general'' CP transformations appear for any space that is already constrained to transform as representation of a symmetry (this can, but does not have to be a flavor symmetry). 
In this case, outer automorphisms such as CP cannot be generalized ``by hand'', but, in fact, have to obey a consistency condition 
that enforce transformation of the fields with a non-trivial representation matrix -- just as is the case for all other symmetry constrained spaces~\cite[Sec.~4.3]{Trautner:2016ezn}.
One is familiar with such ``general'' CP transformations, for example from the transformation of Dirac spinor fields as $\Psi(x)\xmapsto{\mathrm{CP}}\eta_{\mathrm{CP}}\mathcal{C}\Psi^*(\mathcal{P}x)$,
with a free phase $\eta_{\mathrm{CP}}$ and $\mathcal{C}=\mathrm{i}\gamma^0\gamma^2$ in commonly used Weyl or Dirac bases, where everybody is used to the fact that the outer automorphism comes with a non-trivial transformation matrix.}
This was first discussed in the context of left-right symmetric models~\cite{Ecker:1981wv, Ecker:1983hz} and implemented also for the quark~\cite{Gronau:1986xb,Bernabeu:1986fc}
and lepton sectors of the SM~\cite{Branco:1986gr}. 
The requirement of a conserved generalized CP transformation on an otherwise symmetry-unconstrained flavor space will generally induce new linear horizontal symmetries~\cite{Grimus:1987kn}.
Also, generalization may lead to CP transformations of order larger than two which has, for example, been considered in two Higgs~\cite{Maniatis:2007vn, Branco:2011iw} 
and three Higgs doublet models~\cite{Ivanov:2015mwl,Aranda:2016qmp, Ferreira:2017tvy}.
There exist groups (of the so called type~II~B~\cite{Chen:2014tpa}) in which the general CP transformations of the fields are necessarily of higher order, and 
these appear in some examples below. For higher-order CP transformations, necessarily there appear CP eigenstates which are neither CP even nor CP odd (CP eigenvalues $\pm1$), 
but which are CP ``half--odd'' (or even ``$1/2n$--odd'') with eigenvalues $(-1)^{1/2n}$ for a CP transformation of order $2+2n$ $(n\in\mathbbm{N})$~\cite{Ivanov:2018qni}.
All of such possibilities are taken into account in the next chapter where we will introduce a standard form of general CP transformations. This will lead us to
the discussion of how to differentiate between elementary and composite states for CP transformations of all orders and also for more general, non-CP outer automorphisms.

The main goal of the present paper is to point out that not only P and C, or CP, but also the transformation under other outer automorphisms
should be taken into account when looking for hallmark signs of compositeness. Furthermore we do not limit ourselves to composite product states
(such as the fermion anti-fermion bound states) but extend the discussion also to composite non-product states.\footnote{%
Non-product composite states, by default, correspond to entangled states as they cannot be written as a direct product of states.}

Our present discussion will be more general than the previous literature in the following points:
\begin{itemize}
 \item We keep our discussion general and focus on the group theory of any state, going beyond the treatment of non-relativistic bound states; what counts is just the representation under the symmetry groups and their behavior under outer automorphisms.
 \item We do not exclusively focus on product states but also consider composite states which are non-product states. For example, the regular representation and conjugate representation of finite groups. 
 \item We do not limit ourselves to transformations of order two and, hence, will find examples for a transformation behavior of composite states that is more general than just
 multiplication by a phase factor.
 \item As it is crucial for all of these discussions, we clearly distinguish between flavor spaces and symmetry-constrained representation spaces of groups.
 \item The transformation behavior under outer automorphisms is typically fixed by knowing the transformation properties under all symmetries of a theory. For example, the quantum numbers in Eq.~\eqref{eq:PCQNs} are uniquely fixed once the Lorentz group representation of angular momentum and spin are fixed. The automatic fixture of transformation behavior under outer automorphisms by specifying the transformation properties under the symmetry group is not necessarily the case in our discussion. Specifically, we emphasize that states in \textit{identical} representations of all symmetries can transform \textit{differently} under outer automorphisms, depending on whether they are, at heart, composite states or not. 

 A well-known example for states that transform equally under all symmetries but differently only under outer automorphisms are scalar(vector) and their corresponding pseudo-scalar(pseudo-vector) states.
 As mentioned above, this non-trivial transformation behavior (in the absence of elementary states that transform in such a way) can be considered a hallmark of compositeness.
\end{itemize}
We stress that we are not limiting ourselves to any specific physical system here, but investigate the group theoretical laws in general. 
The logic of our discussion is generally applicable to all groups and automorphisms. We focus on finite groups as examples, mostly because 
those are easy to handle and provide a rich zoo of different cases already for sufficiently small groups. 

We expect that our treatment does have applications not only to mesons and baryons but also to their 
more exotic, multi-quark composite cousins (see e.g.~\cite{Gershon:2022xnn} for a review). 
The actual goal of our discussion is to lay the groundwork to further scrutinize the potential compositeness 
of the presently believed-to-be elementary lepton, quark, scalar and vector states of the SM. 
Another, possibly related, area of application are non-product composite ``dressed'' states of the Poincar\'e group~\cite{Kulish:1970ut,Zwanziger:1972sx}
and their transformation behavior under outer automorphisms.

The paper is organized as follows. In chapter~\ref{sec:Standard_Form_CP} we introduce a standard form of general CP transformations in symmetry representation and flavor spaces.
Subsequently, in chapters~\ref{sec:implications} and~\ref{sec:compositeness} we discuss the physical implications of non-standard CP transformations and the lessons one can learn from them about compositeness. Chapter~\ref{sec:examples} is devoted to illustrate the general discussion with instructive and comprehensible examples for product and non-product representations based on the group $\Sigma(72)$. In chapter~\ref{sec:non-CPouts} we extend the discussion to general, non-CP outer automorphisms, exemplified based on the group $D_8$. Finally, in chapter~\ref{sec:continuousgroups} we extend the discussion to semi-simple Lie groups of the type $\SU{N}$ and comment on claims in the recent literature about the existence of a new, non-trivial CP transformation of $\SU{2N}$. Chapter~\ref{sec:continuousgroups} should be sufficiently self contained such that no excessive study of the other chapters is necessary in order to comprehend it. Some extended proofs and technical details are deferred to appendices.
\section{Standard form of general CP transformations}
\label{sec:Standard_Form_CP}
Consider a quantum field theory with a quantum field $\phi(x)$. If the theory is invariant under a symmetry $G$ (this generally includes global, space-time and gauge symmetries), 
$\phi(x)$ will, in general, transform in a representation of this symmetry. In addition, $\phi(x)$ may be subject to an $\Nf$-fold 
repetition, i.e.\ existence of identical copies that we will refer to as flavors. Hence, in a very explicit way one may spell out $\phi^{\alpha}_{a}(x)$ with indices $a=1,..,\dim{\rep{r}}$\,, 
running over the representation space of all symmetries, as well as $\alpha=1,..,\Nf$\,, running over flavor space, respectively. 
The most general \textit{physical} CP transformation\footnote{%
We refer to physical CP as transformations that are tied to the violation of matter anti-matter symmetry, most prominently (but not necessarily) linked to violation of 
baryon or lepton number, see~\cite{Chen:2014tpa} for more details and~\cite{Trautner:2017vlz,Ratz:2019zak} for mini-reviews.}
then corresponds to a mapping
\begin{equation}
\label{eq:general-cp}
\phi(x) \xmapsto{\mathrm{CP}} \mathcal{U}\,\phi^*(\mathcal{P}x)\,,
\end{equation}
where the star denotes complex conjugation, $\mathcal{P}$ denotes the parity transformation of space-time coordinate $x$ (which we will suppress in the following), and $\mathcal{U}$ collectively denotes the possible, generally non-trivial action in symmetry representation and flavor spaces. Importantly, $\mathcal{U}$ cannot, generally, be absorbed by a basis change and we will discuss below when this is the case.

Due to the complex conjugation transformation, CP in general corresponds to automorphisms $u:G\mapsto G$ of the participating groups 
that must act as complex conjugation on their representations.\footnote{%
Such automorphisms are necessarily outer, if any of the involved representations is complex.}
If the group transformation on the field $\phi(x)$ acts as 
\begin{equation}\label{eq:Gaction}
 \phi(x) \xmapsto{G} \rho(g)\,\phi(x)\quad\forall g\in G\,,
\end{equation}
with representation matrices $\rho(g)$, then the representation matrix $\mathcal{U}$ of the most general CP transformation has to obey the consistency condition\footnote{%
The stated form of the consistency condition is specific to complex conjugation automorphisms and linear action \eqref{eq:Gaction} of the symmetry group~\cite{Holthausen:2012dk,Feruglio:2012cw}.
The concept of the consistency condition is universal and can be generalized, for example, to arbitrary automorphisms~\cite{Fallbacher:2015rea,Fallbacher:2015upf,Trautner:2016ezn} or modular group actions, see~\cite{Baur:2019kwi,Novichkov:2019sqv}. We will use a more general form of the consistency condition 
when discussing non-CP outer automorphisms in section~\ref{sec:non-CPouts}.} 
\begin{equation}\label{eq:consistency}
\rho(u(g)) = \mathcal{U}\,\rho(g)^*\,\mathcal{U}^\dagger\quad\forall g\in G\;.
\end{equation}

Due to the direct product structure of symmetry representation space and flavor space one may easily be convinced that $\mathcal{U}$ factorizes as
\begin{equation}
 \mathcal{U}=\mathcal{A}\otimes U\;,
\end{equation}
where $\mathcal{A}$ carries indices $\alpha,\beta,..$ exclusively in flavor space while $U$ carries indices $a,b,..$ exclusively in the symmetry representation space.
Note that, typically, there is more than one symmetry group\footnote{%
Since we include the space-time symmetries within $G$, Lorentz symmetry plus any gauge or global symmetry is enough to qualify for two groups already.} 
in which case $G=G_1\otimes G_2\otimes ..$ and there are mutually orthogonal representation spaces 
such that $U$ may be further factorized into $U=U_{\rep{r}_{G_1}}\otimes U_{\rep{r}_{G_2}}\otimes ..$, where $\rep{r}_{G_i}$ denote the representation under the respective group. Colloquially speaking, each index of $\phi$ has to transform with its own matrix, and if the corresponding spaces are direct product spaces, so will be the according matrices.

For simplicity of the presentation, we will restrict the following discussion to a single finite group $G$ and the representations of it, noting that the generalization is straightforward. 

While $U$ is dictated by the choice of symmetries and representations via the consistency condition, a common perception is that $\mathcal{A}$ merely
corresponds to ``amending the CP transformation by an unphysical basis change in flavor space''. 
However, this view is too naive, and an important point of the present paper is to stress that the form of $\mathcal{A}$ is also constrained by the symmetries of the model in very specific situations which we will elucidate in the subsequent section. Before that let us discuss certain standard forms that can be achieved 
for $\mathcal{A}$ and $U$.

Since the representations under $G$ are, in general, reducible, it makes sense to decompose a field $\phi$ transforming in a
general representation $\rep{r}=\rep{r}_1\oplus\rep{r}_2\oplus..$ into components that transform as irreducible representations (irreps) $\rep{r}_{i=1,..,\Nr}$, represented by matrices~$\rho_{\rep{r}_i}$. If $\phi$ would contain a single field in the irreducible representation $\rep{r}_i$, we may call this $\phi_{\rep{r}_i}$ and CP would map $\phi_{\rep{r}_i}\mapsto U_{\rep{r}_i}\phi_{\rep{r}_i}^*$. The consistency condition needs to be satisfied for each 
irrep individually, i.e.\ the unitary matrices $U_{\rep{r}_i}$ are given as solutions to 
\begin{equation}\label{eq:consistency_irrep}
\rho_{\rep{r}_i}(u(g)) = U_{\rep{r}_i} \rho_{\rep{r}_i}(g)^* U_{\rep{r}_i}^\dagger\quad\forall g\in G\,,\quad\forall\,i\,.
\end{equation}
Once the basis of the irrep $\rho_{\rep{r}_i}$ is fixed, the matrices $U_{\rep{r}_i}$ are uniquely determined by the consistency condition, up to a global phase.\footnote{%
$U_{\rep{r}_i}$ could also differ by multiplication of an element of the group center. However, for unitary representations the multiplication by center elements 
is equivalent to a multiplication by a global phase by Schur's lemma.}
This implies that all present irreducible representations are mapped onto their own complex conjugate representation\footnote{%
As firstly pointed out in~\cite{Chen:2014tpa}, there are groups of the so-called type~I for which no automorphisms $u$ exists that achieves the simultaneous mapping of all irreps onto their own complex conjugate irreps. While interesting in itself, we will not pursue this case here and focus on situations where such a mapping can be achieved. This could be a case for a type~I group and suitably limited number of representations, or simply for an arbitrary number of representations and a group of the so-called type~II.}~\cite{Chen:2014tpa}.

On the contrary, if there appear ${\Nf}_i$ copies (flavors) of fields in an irrep $\rep{r}_i$, we denote them as $\phi^{\alpha=1,..,{\Nf}_i}_{\rep{r}_i}$ and 
the most general CP transformation, Eq.~\eqref{eq:general-cp}, becomes
\begin{align}
 \phi~=~
  \left(\begin{array}{c}\phi_{\rep{r}_1}^{1}\\\vdots\\\phi_{\rep{r}_1}^{{\Nf}_1} \\ \hline 
   \vdots\\ \hline
\phi_{\rep{r}_\Nr}^{1}\\\vdots\\\phi_{\rep{r}_\Nr}^{{\Nf}_{\Nr}}\end{array}\right)
 ~\xmapsto{\mathrm{CP}}~\mathcal{U}\phi^* =&
 \left(\begin{array}{ccc|c|ccc}
 \nwarrow & & \nearrow & & & & \\
 &\!\!\! \mathcal{U}_{\rep{r}_1} \!\!\!&  & & & & \\
 \swarrow & & \searrow & & & & \\
  \hline 
 & & &  \ddots & & &\\
 \hline
 & & & &  \nwarrow  & & \nearrow   \\
 & & & & & \!\!\!\mathcal{U}_{\rep{r}_{{\Nr}}}\!\!\! & \\
 & & & & \swarrow  & & \searrow 
 \end{array}\right)\,
 \left(\begin{array}{c}\phi_{\rep{r}_1}^{1*}\\\vdots\\\phi_{\rep{r}_1}^{{\Nf}_1*} \\ \hline 
   \vdots\\ \hline
\phi_{\rep{r}_{\Nr}}^{1*}\\\vdots\\\phi_{\rep{r}_{\Nr}}^{{\Nf}_{\Nr}*}\end{array}\right)\;.
\label{eq:vdecomposition}
\end{align} 
The matrices $\mathcal{U}_{\rep{r}_i}$ here are \mbox{$(\dim \rep{r}_i\cdot {\Nf}_i)^2$}-dimensional and act on the vectors $\vec\phi_{\rep{r}_i}:=(\phi_{\rep{r}_i}^{1},..,\phi_{\rep{r}_i}^{{\Nf}_i})$.
Due to the orthogonality of symmetry and flavor space, the matrices $\mathcal{U}_{\rep{r}_i}$ factorize as
\begin{equation}\label{eq:acal}
\mathcal{U}_{\rep{r}_i} = \mathcal{A}_{\rep{r}_i} \otimes U_{\rep{r}_i} = \mat{
a_{11}U_{\rep{r}_i} &\dots & a_{1{\Nf}_i} U_{\rep{r}_i}\\ 
\vdots &\ddots &\vdots\\
a_{{\Nf}_i1}U_{\rep{r}_i} & \dots & a_{{\Nf}_i{\Nf}_i} U_{\rep{r}_i}},
\end{equation}
where $\mathcal{A}_{\rep{r}_i}$ is an $({\Nf}_i\times {\Nf}_i)$-dimensional unitary\footnote{%
Since $\mathcal{U}_{\rep{r}_i}$ and $U_{\rep{r}_i}$ are both unitary also $\mathcal{A}_{\rep{r}_i}$ must be unitary, 
$\mathcal{U}_{\rep{r}_i}^\dagger \mathcal{U}_{\rep{r}_i} = \mathcal{A}_{\rep{r}_i}^\dagger\mathcal{A}_{\rep{r}_i} \otimes U_{\rep{r}_i}^\dagger U_{\rep{r}_i} \stackrel{!}{=} \mathbbm{1}\,$.
} 
matrix with components $(\mathcal{A}_{\rep{r}_i})_{\alpha\beta}=a_{\alpha\beta}\in\mathds{C}$.
We will refer to this as the \textit{factorized form} of $\mathcal{U}_{\rep{r}_i}$. 
We give additional motivation for the fact that $\mathcal{U}_{\rep{r}_i}$ can always be factorized in the stated way in Appendix~\ref{sec:Standardform_Derivation}.

It is crucial to note that a basis change in field space $\phi'=\mathcal{X}\phi$ transforms the CP transformation matrix $\mathcal{U}$ as
\begin{equation}
\label{eq:acaltransform}
\mathcal{U}'~=~\mathcal{X}\,\mathcal{U}\,\mathcal{X}^\mathrm{T}. 
\end{equation}
Importantly, there is a transposition and not a Hermitian conjugation on the right-hand side. This is why $\mathcal{U}$ cannot always be absorbed by a basis change. 
In the mathematical literature such a transformation is called a \textit{unitary congruence transformation}~(UCT)~\cite{youla_1961}. 

On the level of irreducible representations, the basis change reads 
\begin{equation}
\mathcal{U}_{\rep{r}_i}'~=~\mathcal{X}_{\rep{r}_i} \mathcal{U}_{\rep{r}_i} \mathcal{X}^\mathrm{T}_{\rep{r}_i}. 
\end{equation}
The factorization into flavor- and symmetry-representation spaces, furthermore, motivates to consider basis changes which themselves factorize as
\begin{equation}
 \mathcal{X}_{\rep{r}_i}~=~X_{\mathcal{A}}\otimes X_U\;.
\end{equation}
This is nothing more than the statement that the flavor space can be rotated independently of the symmetry representation space, 
which is no surprise as they are orthogonal spaces. Hence, each of the rotations $X_{\mathcal{A}}\otimes\mathbbm{1}$ and  $\mathbbm{1}\otimes X_U$ 
acts exclusively on the respective CP transformation matrices $\mathcal{A}_{\rep{r}_i}$ or $U_{\rep{r}_i}$ as a UCT.

Using the available freedom of a UCT, the respective CP transformation matrices can be brought to a standard form~\cite{Ecker:1987qp}(see also~\cite[Ch.2, App.C]{Weinberg:1995mt2}). 
The achievable standard form depends decisively on the symmetry properties of $\mathcal{A}_{\rep{r}_i}$ and $U_{\rep{r}_i}$.
Since the discussion of the standard form is equivalent for $\mathcal{A}_{\rep{r}_i}$ and $U_{\rep{r}_i}$, we will focus on $\mathcal{A}_{\rep{r}_i}$ and rotations by $X_\mathcal{A}$ in the following for concreteness. 
There are three case to be distinguished:
\begin{enumerate}
    \item[1.] $\mathcal{A}_{\rep{r}_i}=\mathcal{A}_{\rep{r}_i}^\mathrm{T}$ is symmetric. Then a UCT
    can be used to rotate $\mathcal{A}_{\rep{r}_i}$ to the identity matrix.
    \item[2.] $\mathcal{A}_{\rep{r}_i}=-\mathcal{A}_{\rep{r}_i}^\mathrm{T}$ is antisymmetric. Then a UCT can be used to rotate $\mathcal{A}_{\rep{r}_i}$ to
    \begin{equation}
    \label{eq:sigma-}
    \mathcal{A}_{\rep{r}_i}=\Sigma_-\oplus\dots\oplus\Sigma_-\,,\quad\text{where}\quad \Sigma_-:= \mat{0&1\\-1&0}\,.
    \end{equation}
    Note that antisymmetry of the unitary $\mathcal{A}_{\rep{r}_i}$ automatically implies it must be of even dimension. 
    \item[3.] $\mathcal{A}_{\rep{r}_i}$ has mixed symmetry, this is the most general case. The most minimal form achievable by UCTs for a general unitary matrix is~\cite{Ecker:1987qp}(see also~\cite[Ch.2, App.C]{Weinberg:1995mt2})
\begin{align}
\label{eq:standard-form}
\mathcal{A}_{\rep{r}_i} &= 
 \mathbbm{1}_{m_i} \oplus \left[\bigoplus_{j=1}^{\ell_i} \mat{\phantom{-}\cos \alpha_{ij} & \sin \alpha_{ij}\\
-\sin\alpha_{ij} & \cos \alpha_{ij}}\right],\quad \text{with}\quad m_i+2\ell_i=n_i\,.
\end{align}
Here $m_i$ is the degeneracy of the eigenvalue $1$ of $\mathcal{A}_{\rep{r}_i}\mathcal{A}_{\rep{r}_i}^*$, while the angles $\alpha_{ij}$ are determined by 
half the arguments of the $\ell_i$ pairwise appearing eigenvalues $\mathrm{e}^{\pm2\I\alpha_{ij}}$ of $\mathcal{A}_{\rep{r}_i}\mathcal{A}_{\rep{r}_i}^{*}$ and can be chosen
in the range $0\leq\alpha_{ij}\leq \pi/2$.\footnote{%
This most general form also contains cases 1. and 2. upon noticing that $\mathcal{A}_{\rep{r}_i}\mathcal{A}_{\rep{r}_i}^*=\pm\mathcal{A}_{\rep{r}_i}\mathcal{A}_{\rep{r}_i}^\dagger=\pm\mathbbm{1}$ for the symmetric/antisymmetric case, 
which has exclusively eigenvalues $\pm1$, respectively.}
\end{enumerate}
A general proof of these statements can be found in~\cite{Ecker:1987qp}. We give an alternative proof based on a more general set of arguments~\cite{youla_1961} in Appendix~\ref{sec:standard-appendix}.

The very same classification in symmetric, antisymmetric or mixed-symmetric can be applied to the matrix $U_{\rep{r}_i}$.
Hence, $U_{\rep{r}_i}$ can be brought to analogous standard forms by the UCT with $X_U$. 

\section{Physical implications}\label{sec:implications}
In this section we will discuss physical implications that arise from non-trivial CP transformation behavior, 
as well as physical circumstances that can enforce such a non-standard transformation behavior. 
In particular, we will see how composite states, such as, for example, product bound states, 
can be distinguished from elementary states by their different CP transformation behavior.
To stay in line with the flow of the paper we focus on a CP outer automorphism here while noting
that the discussion straightforwardly generalizes also to general outer automorphism and we will discuss those in section~\ref{sec:non-CPouts}.

As established in the last section, $\mathcal{U}_{\rep{r}_i}$ cannot, generally, be rotated away by a basis change.
This implies that non-trivial general CP transformations generally have physical consequences.
For example, applying the CP transformation twice one arrives at
\begin{equation}
\label{eq:general-cp2}
\vec\phi_{\rep{r}_i} \xmapsto{(\mathrm{CP})^2} \mathcal{U}_{\rep{r}_i}\,\mathcal{U}_{\rep{r}_i}^*\,\vec\phi_{\rep{r}_i}=\left(\mathcal{A}_{\rep{r}_i}\mathcal{A}_{\rep{r}_i}^*\otimes U_{\rep{r}_i}U_{\rep{r}_i}^*\right)\,\vec\phi_{\rep{r}_i}\;.
\end{equation}
This shows that CP transformations with non-trivial $\mathcal{A}_{\rep{r}_i}$ and/or $U_{\rep{r}_i}$ can enforce new linear symmetries if $\mathcal{U}_{\rep{r}_i}\mathcal{U}_{\rep{r}_i}^*\neq\mathbbm{1}$~\cite{Grimus:1987kn}.\footnote{%
Note that for non-trivial $U_{\rep{r}_i}$ there is the possibility that the CP transformation in the symmetry constrained space closes to an inner automorphism.
That is, $U_{\rep{r}_i}U_{\rep{r}_i}^*=\rho_{\rep{r}_i}(g)\neq\mathbbm{1}$ is equal to an element of the already present symmetry $G$.
Effectively this is the same case as an order $2$ CP transformation and no new linear symmetry would be enforced in the symmetry constrained space.}
For a CP transformation that closes to the identity only after $2n$ applications, the enforced linear symmetry generated 
by $\mathcal{U}_{\rep{r}_i}\mathcal{U}_{\rep{r}_i}^*$ is of order $n$. 
Imposing in flavor space such non-trivial CP transformations of higher order leads to the phenomenologically unprecedented occurrence of CP eigenstates that are 
neither CP even nor CP odd, but CP ``half--odd'' -- for CP transformations of order $4$, or even more exotic 
CP eigenstates for higher-order CP transformations, see e.g.~\cite{Ivanov:2015mwl,Ivanov:2018qni}.

Whether or not a general CP transformation is of higher order depends on the symmetry properties of $\mathcal{U}_{\rep{r}_i}$ under transposition.
The decomposition of $\mathcal{U}_{\rep{r}_i}$ into the factorized form of Eq.~\eqref{eq:acal} clearly shows that 
the transformation behavior of $\mathcal{U}_{\rep{r}_i}=\mathcal{A}_{\rep{r}_i}\otimes U_{\rep{r}_i}$ 
is inherited from the transformation behavior of $\mathcal{A}_{\rep{r}_i}$ and $U_{\rep{r}_i}$. 
For example, for (anti-)symmetric $\mathcal{A}_{\rep{r}_i}$ and/or $U_{\rep{r}_i}$ the transformation
behavior of $\mathcal{U}_{\rep{r}_i}$ is straightforwardly obtained as summarized in Tab.~\ref{tab:decision-table}.

\renewcommand{\arraystretch}{1.2}
\begin{table}[t]
\centering
\begin{tabular}{l|ll}
\hline
Property of $\mathcal{U}_{\rep{r}_i}$ & $U_{\rep{r}_i}$ sym. & $\ur{i}$ antisym.\\
\hline
$\acurlr{i}$ sym. & $\mathcal{U}_{\rep{r}_i}$ sym. &  $\mathcal{U}_{\rep{r}_i}$ antisym. \\
$\acurlr{i}$ antisym. & $\mathcal{U}_{\rep{r}_i}$ antisym. & $\mathcal{U}_{\rep{r}_i}$ sym. \\
\hline
\end{tabular}
\caption{\label{tab:decision-table}%
Decision table for the symmetry properties of matrices $\mathcal{U}_{\rep{r}_i}=\acurlr{i}\otimes\ur{i}$ under transposition, 
depending on the properties of the flavor space dependent part of the CP transformation $\acurlr{i}$ as well as 
on the representation dependent internal part $\ur{i}$.
}
\end{table}
\renewcommand{\arraystretch}{1.0}
A crucial question is whether and how the transformation behavior of $\mathcal{U}_{\rep{r}_i}$ is fixed or subject to a choice. 
Clearly, $U_{\rep{r}_i}$ is subject to constraints from the present symmetries in the form of the consistency condition. 
For example, antisymmetric $U_{\rep{r}_i}$'s are required for certain representations of discrete groups of the so-called type~II~B~\cite{Chen:2014tpa} 
(in fact, the advent of necessarily antisymmetric $U_{\rep{r}_i}$'s for physical CP transformations is the defining property of these groups).
Another example is the transformation of Dirac spinor fields, $\Psi(x)\xmapsto{\mathrm{CP}}\eta_{\mathrm{CP}}\mathcal{C}\Psi^*(\mathcal{P}x)$, where $\mathcal{C}=\mathrm{i}\gamma^0\gamma^2=-\mathcal{C}^\mathrm{T}$ 
is antisymmetric.

By contrast, a common perception is that the form of $\mathcal{A}_{\rep{r}_i}$ would always be unconstrained by the symmetries of a model and rather subject to 
\textit{a model building choice}. One of the main points of the present paper is to stress that this is not true in general. 
Depending on specific physical situation, $\mathcal{A}_{\rep{r}_i}$ may very well be \textit{forced} to obey a certain form \textit{by the symmetries of the model}. 
Specifically, this can be the case if $\phi$ is composite. 
For example, $\phi$ could describe a bound state originating from a direct product of more elementary, constituent fields which 
themselves also transform under $G$. Alternatively, $\phi$ could also be a non-product composite state transforming in a special (reducible) representation of the group such as, for example, 
the regular representation or the conjugate representation which we will use as examples below.

Let us go over the simple example of a product bound state to illustrate our point. 
Consider a theory with two elementary fields $\varphi_{\rep{r}_{k}}$ and $\varphi_{\rep{r}_{\ell}}$ transforming in distinct 
irreps $\rep{r}_{k}$ and $\rep{r}_{\ell}$ of a symmetry group $G$. Under a CP outer automorphism the elementary fields transform as $\varphi_{\rep{r}_{k,\ell}}\xmapsto{\mathrm{CP}} U_{\rep{r}_{k,\ell}}\varphi_{\rep{r}_{k,\ell}}^{*}$,
where $U_{\rep{r}_{k,\ell}}$ are uniquely determined (up to a global phase) by the corresponding consistency conditions. 
Now consider the direct product state $\varphi_{\rep{r}_{k}}\otimes\varphi_{\rep{r}_{\ell}}$. Such a state could, for example, 
arise as a ``meson'' bound state if we switch on an additional confining interaction between $\varphi_{\rep{r}_{k}}$ and $\varphi_{\rep{r}_{\ell}}$.
Under CP the composite state transforms as 
\begin{equation}
\varphi_{\rep{r}_{k}}\otimes\varphi_{\rep{r}_\ell} \xmapsto{\mathrm{CP}} \left(U_{\rep{r}_k}\varphi_{\rep{r}_k}^*\right)\otimes \left(U_{\rep{r}_\ell}\varphi_{\rep{r}_\ell}^*\right) 
= \left(U_{\rep{r}_k}\otimes U_{\rep{r}_\ell}\right)\left(\varphi_{\rep{r}_k}^*\otimes\varphi_{\rep{r}_\ell}^*\right)\,=:\,\mathcal{U}\,\left(\varphi_{\rep{r}_k}^*\otimes\varphi_{\rep{r}_\ell}^*\right).
\label{eq:ProductState_Ex}
\end{equation}
Clearly, the overall CP transformation matrix $\mathcal{U}$ of the product state is uniquely fixed by the transformation of the constituent states. 
Furthermore, one can perform a basis change to realize covariantly transforming composite states 
\begin{equation}
 \phi_{\rep{r}_i}\subset\varphi_{\rep{r}_{k}}\otimes\varphi_{\rep{r}_\ell}\;,
\end{equation}
which themselves transform as irreps under $G$. For this we perform a basis change with a matrix $S$ such that
\begin{equation}
\begin{split}
S\left(\rho_{\rep{r}_k}\otimes\rho_{\rep{r}_\ell}\right)S^\dagger &= \bigoplus_{i=1}^{{\Nr}}\left(\rho_{\rep{r}_i}\right)^{\oplus {\Nf}_i}\,,\\
S\left(\varphi_{\rep{r}_k}\otimes\varphi_{\rep{r}_\ell}\right) &= \left(\phi_{\rep{r}_1}^1\oplus\dots\oplus\phi_{\rep{r}_1}^{{\Nf}_1}\right) \oplus \dots \oplus \left(\phi_{\rep{r}_{{\Nr}}}^1\oplus \dots \oplus \phi_{\rep{r}_{{\Nr}}}^{{\Nf}_{\Nr}}\right)\,.
\label{eq:ProductState_Ex2}
\end{split}
\end{equation}
Here, $\Nr$ is the number of different irreps in the direct product representation and ${\Nf}_i$ denotes the multiplicity 
(number of flavors) of fields in each irrep $\rep{r}_{i}$.
Accordingly, $\mathcal{U}$ transforms as $\mathcal{U}^{\prime} = S\,\mathcal{U}\,S^\mathrm{T}$ and acquires a block diagonal form
with blocks $\mathcal{U}^{\prime}_{\rep{r}_i}=\mathcal{A}_{\rep{r}_i}\otimes U_{\rep{r}_i}$ that individually factorize as in Eq.~\eqref{eq:acal}.
The transformation property of $\mathcal{U}$ under transposition, and consequently the transformation properties of the contained blocks $\mathcal{U}_{\rep{r}_i}$ under transposition, are preserved under the 
basis transformation. Since the CP transformation matrices $U_{\rep{r}_i}$ of the sub-block irreps and specifically their transformation under transposition are fixed by group theory, 
it follows that also the transformation properties and, hence, standard forms of the matrices $\mathcal{A}_{\rep{r}_i}$ are uniquely fixed. 
Consequently, in this example there is \textit{no} freedom in choosing the symmetry properties of the flavor space transformation matrices $\mathcal{A}_{\rep{r}_i}$, 
but they are entirely fixed by group theory.

For example, if $U_{\rep{r}_{k}}$ would be symmetric and $U_{\rep{r}_{\ell}}$ antisymmetric then $\mathcal{U}_{\rep{r}_i}$ must be antisymmetric as well.
In case $\rep{r}_i$ would be a representation with symmetric $U_{\rep{r}_i}$, it follows from Tab.~\ref{tab:decision-table} that $\mathcal{A}_{\rep{r}_i}$
\textit{must} be antisymmetric. In view of the achievable standard form for such a transformation matrix, see Eq.~\eqref{eq:sigma-}, this 
would imply that $(i)$ $\phi_{\rep{r}_i}$ contains an even number of copies of composite states transforming as $\rep{r}_i$ of $G$, which $(ii)$ get
pairwisely interchanged under the CP transformation. 
This implies that these composite states in $\phi_{\rep{r}_i}$ would transform differently under CP than an elementary state in representation $\rep{r}_i$, 
even though both states are indistinguishable by their transformation behavior under $G$ alone because they transform in exactly the same 
representation. 
This illustrates how CP transformations and specifically the symmetry properties of the flavor transformation matrix 
$\mathcal{A}_{\rep{r}_i}$ can be used to study the compositeness of a given state.

Analogous situations arise for non-product composite states. In this case $\mathcal{U}$ cannot be simply computed as 
a direct product of $U_{\rep{r}_i}$'s but has to be computed differently. The rest of the discussion remains the same. 

It is worth discussing the special case of so-called type~II~A groups~\cite{Chen:2014tpa}. For those groups, by definition, 
there exist a class-inverting outer automorphism with symmetric representation matrices for all irreps, $U_{\rep{r}_i}^\mathrm{T}=U_{\rep{r}_i}~\forall~i$ (following~\cite{Chen:2014tpa}, we call this a Bickerstaff-Damhus automorphism~\cite{Bickerstaff:1985jc}).
This implies that the according CP transformation matrix $\mathcal{U}$ of all possible product states will always 
be symmetric under transposition. Hence, also all flavor space transformation matrices $\mathcal{A}_{\rep{r}_i}$ have 
to be symmetric under transposition. Thus, a basis can be found in which the CP transformation is trivial for elementary
and composite product states alike. We stress that the same discussion applies to $\SU{N}$ groups, because also for them
the class inverting outer automorphism used to define CP comes with symmetric transformation matrices~\cite{Grimus:1995zi} (see also our discussion in 
section~\ref{sec:continuousgroups}). We stress that this argument does not apply to \textit{non}-product composite states.
For the case of non-product composite states we could not find an argument to generally exclude the presence of an anti- or mixed-symmetric $\mathcal{U}$.
Given such a state exists, the anti- or mixed symmetric $\mathcal{U}$ together with the symmetric $U_{\rep{r}_i}$'s would imply a necessarily non-trivial transformation in flavor space.

\section{Telling Compositeness}
\label{sec:compositeness}
The discussion of the preceding section gives us, in principle, the tools to decide whether an observed state is at heart elementary or composite. 
The distinction arises because states in one and the same irrep $\rep{r}_i$ of a symmetry group $G$ can transform differently under the outer automorphisms
of the symmetry depending on whether they are elementary or composite. While the transformation matrix $U_{\rep{r_i}}$ of an elementary state is fixed by the consistency condition(s), composite states can, in addition, pick up a non-trivial transformation matrix 
$\mathcal{A}_{\rep{r}_i}$. In some cases this difference only corresponds to a relative phase in the transformations of composite as compared to elementary states. In other cases it implies a non-trivial exchange of composite states in flavor space. 

Clearly, a distinction of elementary and composite states in such a way is based on the quantum numbers of states with respect to the outer automorphism.
This distinction, hence, could be made from the long distance without explicitly resolving the potential constituents of a composite state, say, 
in a short distance scattering experiment. 

An important caveat is that the transformation behavior under outer automorphisms and compositeness is not in a one-to-one relation for the following reason. 
Assume we arrive at an experimental situation where we had determined a bunch of candidate states transforming in irreducible representations $\rep{r}_i$ and want to determine if these are composite or elementary states. Probing the states' transformation behavior under an outer automorphism, one possible result is that they transform equivalently to composite states -- which would give strong hints towards their compositeness. However, it could not be excluded that the states are indeed a collection of elementary states that transform under $\rep{r}_i$ but are accidentally (or conspiratively) mixed in such a way that they transform 
under the outer automorphism exactly \textit{as if they were} composite states. This could always be engineered by imposing a non-trivial transformation matrix $\mathcal{A}_{\rep{r}_i}$ on a set of elementary states, where $\mathcal{A}_{\rep{r}_i}$ is such that it corresponds to the transformation matrix of a composite state. On the contrary, if we were to find states which are \textit{not} transforming like composites, their composite nature could immediately be excluded. 
If all possible compositions can be excluded (different constituents may lead to different transformation behavior under outer automorphisms), 
this would prove the elementary nature of a state. 

This shows that a positive proof of compositeness may not, in general, be possible, while the hypothesis of compositeness could unequivocally be rejected.
Note that nothing in our discussion depends on the bound state being composed out of only two constituent fields. Generalizations to product states with more than two factors are straightforward. 

It is important to stress that direct product states are not the only example for situations in which $\mathcal{U}$ and, therefore, the subordinate $\mathcal{A}_{\rep{r}_i}$'s can be uniquely fixed. Extending the whole discussion to continuous groups, examples for physically relevant non-product states are given by gauge invariant ``dressed'' states~\cite{Kulish:1970ut} and in particular composite states of the Poincar\'e group that have both electric and magnetic charges~\cite{Zwanziger:1972sx}, see~\cite{Csaki:2020inw,Csaki:2020yei} for an up-to-date discussion of Zwanziger's pairwise little group, as well as~\cite{Lippstreu:2021avq} for a discussion that involves an additional boost subgroup. While it is beyond the scope of this work, we think that it would be extremely interesting to investigate the potential non-standard CP transformation behavior of such states, in particular, in the context of CP (time-reversal) violation generated by these states directly.

In the next section we will give explicit examples for product states and also non-product states to illustrate our discussion.
Explicit examples for non-product representations that we will treat are special reducible representations of finite groups, 
such as the regular representation and the conjugate representation.
\section{Examples}\label{sec:examples}
In the following, we will discuss examples for the CP transformation behavior of product and non-product representations.
Most interesting in this context are groups of the type~II~B, because they, by definition, feature CP-type automorphisms 
which transform (at least) some irreps with antisymmetric matrices. 
We will discuss the smallest type~II~B group $\Sigma(72)\cong\SG{72,41}$.\footnote{%
We refer to groups using the \texttt{SmallGroupID} of the GAP discrete algebra program~\cite{GAP4}.}
Details of this group are deferred to Appendix~\ref{app:Sig72SG144}.
We note that $\Sigma(72)$ is special because it is ambivalent in the mathematical sense,
meaning that each element is conjugate to its inverse. This implies that inner automorphisms are class-inverting, 
and also CP transformations can be defined based on inner, or even the trivial (identity) automorphisms. 
We mention this special property here just to stress that it does \textit{not} affect our discussion, which 
equally well applies to non-ambivalent groups and non-trivial class-inverting automorphisms.
For example, we have checked this explicitly for the next-to-smallest type~II~B group \SG{144,120},
which is non-ambivalent and features complex irreps. However, since our points are sufficiently
well demonstrated based on $\Sigma(72)$ we focus on this group for brevity.
\subsection{Product representations}\label{subsec:PR}

The group $\Sigma(72)$ can be generated by two elements $P$ and $M$ and has four real one-dimensional, one real eight-dimensional and a pseudo-real two-dimensional irrep, see Appendix~\ref{app:Sig72} for details. 
For $\Sigma(72)$, the identity automorphism is class-inverting and involutory.
Hence, it can be used to define a CP transformation that acts on the irreps as 
\begin{equation}
(M,P)\xmapsto{\mathrm{id}} (M,P)\quad\curvearrowright\quad\mathbf{1}_i\mapsto\mathbf{1}_i^*\,,\quad\mathbf{2}\mapsto U_{\rep{2}}\:\mathbf{2}^*\,,\quad\mathbf{8}\mapsto U_{\rep{8}}\:\mathbf{8}^*\,,
\label{eq:CP72MapRule}
\end{equation}
where the consistency condition~\eqref{eq:consistency} solved in our basis, see Eqs.~\eqref{eq:S72gens2D} and \eqref{eq:S72gens8D}, reveals that
\begin{equation}
  U_{\rep{2}} ~=~ \Sigma_- ~=~ \begin{pmatrix}0 & 1\\ -1 & 0 \end{pmatrix}\;,\quad\text{and}\quad  U_{{\rep{8}}} ~=~ \mathbbm{1}_8\;.
\end{equation}
Hence, our chosen basis realizes the standard form of general CP transformations of section~\ref{sec:Standard_Form_CP}. 

Consider now a state $\phi_{\rep{8}\otimes\rep{2}}$ transforming under $\Sigma(72)$ in the direct product representation 
\begin{equation}
\rho_{\rep{8}\otimes\rep{2}}\equiv \rho_\mathbf{8}\otimes \rho_\mathbf{2}\;.
\end{equation}
$\phi_{\rep{8}\otimes\rep{2}}$ can be decomposed into irreducible representations,
\begin{equation}
\phi_{\rep{8}\otimes\rep{2}}\cong\phi^{(1)}_{\rep{8}}\oplus \phi^{(2)}_{\rep{8}}\;,
\end{equation}
transforming with matrices $\rho_\mathbf{8}\oplus \rho_\mathbf{8}$.
The basis change matrix $S$ that realizes this block-diagonalization, 
\begin{equation}
S\left[\rho_\mathbf{8}(g)\otimes\rho_\mathbf{2}(g)\right]S^\dagger = \rho_\mathbf{8}(g)\oplus\rho_\mathbf{8}(g)
\quad\forall g\,,
\end{equation}
is straightforwardly obtained from the Clebsch-Gordan coefficients of the group, or can conveniently be calculated using the algorithm
for simultaneous block diagonalization (SBD) of~\cite{Bischer:2020nwl}.

Under the CP transformation Eq.~\eqref{eq:CP72MapRule}, $\phi_{\rep{8}\otimes\rep{2}}\mapsto \mathcal{U}\phi_{\rep{8}\otimes\rep{2}}^{*}$, where the transformation matrix $\mathcal{U}$ can be obtained from the direct product
\begin{equation}
\mathcal{U}=U_\mathbf{8} \otimes U_\mathbf{2} = \mathbbm{1}_8 \otimes \Sigma_- = \bigoplus_{n=1}^{8} \Sigma_-\,.
\end{equation}
Clearly, $\mathcal{U}$ is antisymmetric. Transforming $\mathcal{U}$ from the direct-product basis to the block diagonal basis yields
\begin{equation}
\mathcal{U}^{\prime} = S\,\mathcal{U}\,S^\mathrm{T}=\Sigma_-\otimes \mathbbm{1}_8 = \mat{\mathbf{0}_8&\mathbbm{1}_8\\ -\mathbbm{1}_8 & \mathbf{0}_8}.
\end{equation}
This shows how $\mathcal{U}^{\prime}$ decomposes into the factorized form $\mathcal{A}\otimes U_{\rep{8}}$, where $U_{\rep{8}}=\mathbbm{1}_8$ and 
the resulting $\mathcal{A}=\Sigma_-$.

We observe that the CP transformation enforces the non-trivial interchange of the two composite $\rep{8}$-plet states,
\begin{equation}
\phi^{(1)}_{\rep{8}}\mapsto\left(\phi^{(2)}_{\rep{8}}\right)^* \qquad \text{and} \qquad 
\phi^{(2)}_{\rep{8}}\mapsto-\left(\phi^{(1)}_{\rep{8}}\right)^*\;. 
\end{equation}
The CP transformation here acts differently on $\rep{8}$-plet ``mesons'' $\phi^{(1),(2)}_{\rep{8}}$ (composed out of $\rep{8}\otimes\rep{2}$)
than it would act on a ``constituent'' $\rep{8}$-plet, which transforms under CP according to Eq.~\eqref{eq:CP72MapRule}.
This shows how the composite and elementary states can be distinguished by their transformation behavior under the outer automorphism.

\subsection{Non-product representations}\label{subsec:NPR}
Next to product representations, another case of interest are non-product representations. These are reducible representations of a matrix group that cannot be written 
in terms of direct products of irreps. Two examples for non-product representations that exist for all finite groups are the regular and the conjugate representation. 
We will study these in the following.

\paragraph{1. Regular representation of $\boldsymbol{\Sigma(72)}$.}
The (left-)regular representation is given by the action of a group on itself by (left-)multiplication\footnote{%
For non-Abelian groups there is also a right-regular representation that should not concern us in the following.}, i.e.\ $\mathrm{reg}_g(h):=gh$ with $g,h\in G$.
Assigning each group element a canonical basis vector in $\mathbb{R}^n$, where $n:=|G|$ is the order of the group, 
the regular representation of an element $g$ acts on this space as an $n\times n$ permutation matrix. 
We call this the permutation basis of the regular representation. In this basis, the action of an (outer) automorphism on the regular representation 
is straightforwardly obtained from its permutation action on the generators. 

The matrix representations of the generators for the regular representation of $\Sigma(72)$ in the permutation basis are stated in App.~\ref{app:Sig72}. Since the automorphism used to define CP here is the identity automorphism, see~\eqref{eq:CP72MapRule}, its action
in the permutation basis is given by $U_\mathrm{reg}=\mathbbm{1}_{72}$. It is straightforward to see that $U_\mathrm{reg}$ indeed solves the consistency condition, Eq.~\eqref{eq:consistency}, as all representation matrices are real in the permutation basis.

The decomposition of the regular representation into irreps contains all irreps of the group
with a multiplicity being equal to their dimensionality, see e.g.~\cite{Ramond:2010zz}.
This block diagonalization can be explicitly realized by a matrix $S$
\begin{align}\nonumber
    S\,\rho_\mathrm{reg}(M)\,S^{\dagger}&=\rho_{\mathbf{1}_0}(M)\oplus\rho_{\mathbf{1}_1}(M)\oplus\rho_{\mathbf{1}_2}(M)\oplus\rho_{\mathbf{1}_{3}}(M)\oplus 2\cdot\rho_{\mathbf{2}}(M)\oplus 8\cdot\rho_{\mathbf{8}}(M)\,,&\\
    S\,\rho_\mathrm{reg}(P)\,S^{\dagger}&=\rho_{\mathbf{1}_0}(P)\oplus\rho_{\mathbf{1}_1}(P)\oplus\rho_{\mathbf{1}_2}(P)\oplus\rho_{\mathbf{1}_3}(P)\oplus 2\cdot\rho_{\mathbf{2}}(P) \oplus 8\cdot\rho_{\mathbf{8}}(P)\,.&
\end{align}
The transfer matrix $S$ needs to be known explicitly in order to find the action of the CP transformation, $U_\mathrm{reg}$, in the block diagonal basis.
Finding the transfer matrix for large representations is, in general, a tedious task. This can be massively speed up by employing the SBD algorithm~\cite{Bischer:2020nwl}.
Using this to compute $S$, we find in the block diagonal basis
\begin{equation}
U^{\prime}_{\mathrm{reg}}=S\,U_\mathrm{reg}\,S^\mathrm{T}=\mat{\mathbbm{1}_4 \\ & \mathcal{U}_\mathbf{2} \\ &&\mathbbm{1}_{64}},
\end{equation}
with
\begin{equation}\label{eq:U2}
\mathcal{U}_{\rep{2}}=\Sigma_-\otimes\Sigma_- 
=
\left(
\begin{array}{cccc}
 0 & 0 & 0 & 1 \\
 0 & 0 & -1 & 0 \\
 0 & -1 & 0 & 0 \\
 1 & 0 & 0 & 0 \\
\end{array}
\right).
\end{equation}
We see that composite states in the regular representation transforming in the one-dimensional or eight-dimensional representations 
transform under CP like the elementary states in Eq.~\eqref{eq:CP72MapRule}. Contrary to that, the two composite doublet states in the regular representation mix under CP transformations via the $4\times4$-matrix $\mathcal{U}_\mathbf{2}$. In terms of the factorized form \eqref{eq:acal} we have $\mathcal{A}_{\rep{2}}=\Sigma_-$ and $U_\mathbf{2}=\Sigma_-$.
Crucially, note that it is \textit{not} possible to find a basis in such a way that $\mathcal{U}_\mathbf{2}$ and the group
representation matrices are simultaneously block diagonalized. The fact that the CP transformation matrix of the regular representation $U_\mathrm{reg}$ is symmetric 
implies that also $\mathcal{U}_\mathbf{2}$ must be symmetric. Hence, given that the action on the irreps $U_\mathbf{2}$ is antisymmetric (as can be inferred from the twisted FSI), 
the only possibility for $\mathcal{A}_{\rep{2}}$ is to be antisymmetric as well. Consequently, mixing of the two composite doublet states of the regular representation, 
say $\phi_{\rep{2}}^{(1)}$ and $\phi_{\rep{2}}^{(2)}$, under CP is the only possibility. While elementary doublets would transform as in Eq.~\eqref{eq:CP72MapRule},
doublets in the regular representation are enforced to pick up a non-trivial transformation in flavor space, transforming as
\begin{equation}\label{eq:ASdoublets}
\phi^{(1)}_{\rep{2}}\mapsto U_{\rep{2}}\left(\phi^{(2)}_{\rep{2}}\right)^* \qquad \text{and} \qquad 
\phi^{(2)}_{\rep{2}}\mapsto-U_{\rep{2}}\left(\phi^{(1)}_{\rep{2}}\right)^*\;.
\end{equation}
In this example, we find that the doublets contained in the regular representation transform differently than elementary doublets,
in the sense that they are necessarily non-trivially permuted under the CP transformation. 
This would allow one to discriminate composite states of the regular representation from their elementary counterparts.

\paragraph{2. Conjugate representation of $\boldsymbol{\Sigma(72)}$.}
As a second example for non-product representations we investigate the conjugate representation of $\Sigma(72)$. 
The conjugate representation is given by the action of a group on itself by conjugation, i.e.\ $\mathrm{conj}_g(h):=ghg^{-1}$.  
Again, we start in the permutation basis where each element $g$ is associated to a $n\times n$ permutation matrix 
acting on the canonical vectors in $\mathbb{R}^n$ corresponding to the permutation of group elements induced by 
$\mathrm{conj}_g(h)$. Since the CP transformation is based on the trivial automorphism, it 
is in the permutation basis again represented by a matrix $U_\mathrm{conj}=\mathbbm{1}_{72}$.
The permutation matrices corresponding to the $\Sigma(72)$ generators $M$ and $P$ in the conjugate 
representation can be found in \autoref{app:Sig72}. The conjugate representation decomposes into 
\begin{align}
    \rho_\mathrm{conj}~\cong~3\cdot\rho_{\mathbf{1}_0}\oplus 3\cdot\rho_{\mathbf{1}_1}\oplus 3\cdot\rho_{\mathbf{1}_2}\oplus 3\cdot\rho_{\mathbf{1}_3}\oplus 2\cdot\rho_{\mathbf{2}}\oplus 7\cdot\rho_{\mathbf{8}}\;.
\end{align}
Again, we use the SBD algorithm~\cite{Bischer:2020nwl} to derive a matrix $S$ of a basis change that explicitly realizes this block diagonalization. 
Applying this basis change with $S$ to the CP automorphism reveals the transformation behavior of the irreducible representations under CP,
\begin{align}
 U^{\prime}_{\mathrm{conj}}~=~S\,U_{\mathrm{conj}}\,S^\mathrm{T}~=~
 \mat{
 \mathbbm{1}_{12} &0&0 \\ 
 0& \mathcal{U}_{\rep{2}}&0 \\ 
 0&0&\mathbbm{1}_{56}}\;,
\end{align}
where $\mathcal{U}_{\rep{2}}=\Sigma_-\otimes\Sigma_-$ is the same as in Eq.~\eqref{eq:U2}. Hence, states in singlet and octet irreps
of the conjugate representation are mapped onto their own complex conjugates, like their elementary counterparts in Eq.~\eqref{eq:CP72MapRule}.
On the contrary, states in the two copies of the doublet irrep transform non-trivially, and have to be non-trivially interchanged, just as in Eq.~\eqref{eq:ASdoublets}.

This shows that, also for the conjugate representation, there is a non-trivial transformation behavior enforced for certain composite states, 
which would allow one to discriminate them from their elementary counterparts.
\section{Non-CP outer automorphisms}\label{sec:non-CPouts}
Up to now we have limited our discussion to class-inverting involutory (outer) automorphisms, i.e. automorphisms that can be used to define 
proper physical CP transformations. However, these are just particular examples of automorphisms. In the following we generalize our discussion
to generic outer automorphisms. Again, we will focus on finite groups while noting that the extension to other groups is straightforward.
As an example, in \autoref{sec:D8} we will discuss the dihedral group $D_8$ which has a non-trivial $\mathbbm{Z}_2$ outer automorphism which is not class-inverting.
\subsection{Transformation of irreducible representations}
While an automorphism $u$ that can be used to define a proper physical CP transformation has to map irreps to their respective complex conjugate irrep, 
\begin{equation}
\rep{r}_i\;{\xmapsto{~u~}}\;\overline{\rep{r}}_i\;{\xmapsto{~u~}}\;\rep{r}_i\;,
\end{equation}
general outer automorphisms act on the irreps as a mapping of different irreps of equal dimension onto each other,
\begin{equation}\label{eq:out_cycle}
\rep{r}_1\;{\xmapsto{~u~}}\;\rep{r}_2\;{\xmapsto{~u~}}\;\dots\;{\xmapsto{~u~}}\;
\rep{r}_{n}\;{\xmapsto{~u~}}\;\rep{r}_1\;.
\end{equation}
Depending on the details of the group, this may include the possibility of involving multiple irreps and transformations $u$ of arbitrary order.
Focusing on one specific outer automorphism $u:G\mapsto G$, we can sort the irreps $\rep{r}_i$ of $G$ into disjoint sets.
Within each set, $u$ acts as a cyclic permutation of irreps and we will focus on one such set in the following. 
If the transformation $u$ involves a mapping of representations $\rep{r}_i\mapsto\rep{r}_j$, with representation matrices $\rho_{\rep{r}_i}$ and $\rho_{\rep{r}_j}$, 
this implies the most general form of the consistency condition~\cite{Trautner:2016ezn,Trautner:2017vlz}
\begin{equation}
 \rho_{\rep{r}_i}(u(g)) ~=~ U_{\rep{r}_i,\rep{r}_j} \,\rho_{\rep{r}_j}(g)\,U_{\rep{r}_i,\rep{r}_j}^\dagger\,\qquad\forall g\in G,\quad\forall i,j.\\
\end{equation}
The equation has solutions for $U_{\rep{r}_i,\rep{r}_j}$ if and only if $u$ is a valid automorphism of the group 
that connects $\rho_{\rep{r}_i}$ to $\rho_{\rep{r}_j}$.
Note that it is possible to chose bases for $\rho_{\rep{r}_{i}}$ and $\rho_{\rep{r}_{j}}$ such that $U_{\rep{r}_i,\rep{r}_j}$ 
is rotated away. Such a basis may be called a ``$u$-eigenbasis'' for the irreps $\rep{r}_i$ and $\rep{r}_j$.
In turn, if the bases for $\rho_{\rep{r}_{i}}$ and $\rho_{\rep{r}_{j}}$ are fixed from elsewhere, this also fixes the 
representation matrix $U_{\rep{r}_i,\rep{r}_j}$ of an outer automorphism $u$ up to a global phase. 

It is important to note that in the context of a concrete physical model, it makes only sense to consider transformations for which all irreps in the cycle \eqref{eq:out_cycle} are 
part of the model. In the case of CP transformations this is automatically the case because both irreps and their complex conjugate irreps must be present 
upon requiring a real valued Lagrangian. In all other cases, however, there is the possibility of breaking an outer automorphism \textit{explicitly and maximally}, i.e.\ by simply not having one or multiple of the target representations as part of the model. In this case the outer automorphism would not map the theory to itself, but to some other theory. 
Most prominently this is the case, for example, for the SM parity and charge conjugation 
transformations.

We will in the following assume that all irreps in \eqref{eq:out_cycle} are present in the model. In this case we can 
collect the respective fields in a reducible representation $\phi=\phir{1} \oplus\dots\oplus \phir{n}$ and write the transformation under $u$ 
as simple block matrix multiplication 
\begin{equation}\label{eq:trafo_general}
\phi(x) \equiv \mat{\phir{1}(x)\\\phir{2}(x)\\\vdots\\\phir{n}(x)} \;\xmapsto{~u~}\;
\mat{0 & U_{\rep{r}_1,\rep{r}_2} & 0 & 0 &\dots \\ 0 & 0 & U_{\rep{r}_2,\rep{r}_3} & 0 & \dots \\
\vdots & \vdots & \vdots & \ddots &   \\
U_{\rep{r}_{n},\rep{r}_1} & 0 & 0 & 0 &\dots
}
\mat{\phir{1}(x)\\\phir{2}(x)\\\vdots\\\phir{n}(x)}.
\end{equation}
For clarity we have restored the space-time arguments in this equation.
The representation matrices then must fulfill the interlinked consistency conditions
\begin{align}\label{eq:consistency-general}\notag
\rho_{\rep{r}_1}(u(g)) ~&=~ U_{\rep{r}_1,\rep{r}_2} \,\rho_{\rep{r}_2}(g)\, U_{\rep{r}_1,\rep{r}_2}^\dagger\,\qquad\forall g\in G\,,&\\\notag
\rho_{\rep{r}_2}(u(g)) ~&=~ U_{\rep{r}_2,\rep{r}_3} \,\rho_{\rep{r}_3}(g)\, U_{\rep{r}_2,\rep{r}_3}^\dagger\,\qquad\forall g\in G\,,&\\\notag
&\ \,\vdots&\\
\rho_{\rep{r}_{n}}(u(g)) ~&=~ U_{\rep{r}_n,\rep{r}_1} \,\rho_{\rep{r}_1}(g)\, U_{\rep{r}_n,\rep{r}_1}^\dagger\,\qquad\forall g\in G\,.&
\end{align}
For the special case that the corresponding automorphism $u$ is of order $n$, i.e.\ of the order that corresponds to the number of representations in the set $\left\{\rep{r}_1,\dots,\rep{r}_n\right\}$ mapped onto each other under the cycle \eqref{eq:out_cycle}, 
we can furthermore show that one can always achieve a ``$u$-eigenbasis''. 
For this, note that upon applying the consistency condition to $u^n(g)=g$ and successively plugging in the interlinked consistency conditions one obtains 
\begin{equation}
\begin{split}
\rhor{1}(u^n(g)) ~&=~ U_{\rep{r}_1,\rep{r}_2}\,\rhor{2}(u^{n-1}(g))\,U_{\rep{r}_1,\rep{r}_2}^\dagger=\dots =\\
~&=~
\left[U_{\rep{r}_1,\rep{r}_2}\,U_{\rep{r}_2,\rep{r}_3}\dots U_{\rep{r}_{n},\rep{r}_1}\right]\,\rhor{1}(g)\,\left[U_{\rep{r}_1,\rep{r}_2}\,U_{\rep{r}_2,\rep{r}_3}\dots U_{\rep{r}_{n},\rep{r}_1}\right]^\dagger\,\qquad\forall g\in G\,.
\end{split}
\end{equation}
By the assumption of $u$ being of the order $n$, and using Schur's lemma as well as the unitarity of all $U_{\rep{r}_i,\rep{r}_j}$'s, one furthermore finds 
\begin{equation}\label{eq:schur-property}
U_{\rep{r}_1,\rep{r}_2}\,U_{\rep{r}_2,\rep{r}_3}\dots U_{\rep{r}_n,\rep{r}_1}~=~\mathrm{e}^{\I\alpha}\cdot\mathbbm{1}\,, \qquad\mathrm{with}\qquad\alpha\in\mathbbm{R}\;.
\end{equation}
The phase can easily be removed by fixing one of the free phases of any of the $U_{\rep{r}_i,\rep{r}_j}$.
Hence, one may perform basis rotations for all $1<k\leq n$,
\begin{equation}\label{eq:out-eigenbasis}
\rhor{k}(g)':=\left[U_{\rep{r}_1,\rep{r}_2}\,U_{\rep{r}_2,\rep{r}_3}\dots U_{\rep{r}_{k-1},\rep{r}_k}\right]\,\rhor{k}(g)\,\left[U_{\rep{r}_1,\rep{r}_2}\,U_{\rep{r}_2,\rep{r}_3}\dots U_{\rep{r}_{k-1},\rep{r}_k}\right]^\dagger
\end{equation}
which successively removes all $U_{\rep{r}_i,\rep{r}_j}$ from the interlinked consistency conditions, Eq.~\eqref{eq:consistency-general},
upon using the property \eqref{eq:schur-property} of the transformation matrices in the very last of the conditions.
Despite always achievable, we note that working in a ``$u$-eigenbasis'' is typically of limited practicability. This may be the case if 
more than one automorphism is under consideration, or other terms of the Lagrangian such as kinetic or interaction terms should be kept in a 
specific form that does not permit additional basis rotations. 

In the more general case where the number of representations in a cycle~\eqref{eq:out_cycle} is not commensurate with the order of the automorphism
the ``best'' basis that can be achieved is to remove all besides the last transformation matrix in \eqref{eq:consistency-general} as well as the diagonalization 
of the remaining matrix~\eqref{eq:schur-property}. Note that this last and most general case also applies to the special case where the outer 
automorphism maps an irrep to itself, $u:~\rep{r}_i\mapsto\rep{r}_i$, in which case the best that can be achieved for the associated transformation matrix 
is diagonalization.

At this point it may be instructive to reproduce the special case of class-inverting outer automorphisms (used for CP transformations) 
for which $\rep{r}_i\mapsto\bar{\rep{r}}_i\mapsto\rep{r}_i$.
Hence, there are two consistency conditions for these two irreps,
\begin{align}\label{eq:consistency_exampleCP}\notag
\rho_{\rep{r}_1}(u(g)) &= U_{\rep{r}_1,{\overline{\rep{r}}_1}} \,\rho_{\overline{\rep{r}}_1}(g) U_{\rep{r}_1,{\overline{\rep{r}}_1}}^\dagger\,,&\\
\rho_{\overline{\rep{r}}_1}(u(g)) &= U_{{\overline{\rep{r}}_1},\rep{r}_1} \,\rho_{\rep{r}_1}(g) U_{{\overline{\rep{r}}_1},\rep{r}_1}^\dagger\,.&
\end{align}
CP is special in the sense that we usually require to work in a basis for $\rho_{\overline{\rep{r}}_1}$ such that $\rho_{\overline{\rep{r}}_1}(g)=\rhor{1}^*(g)$ and $\phi_{\overline{\rep{r}}_1}=\phir{1}^*$.\footnote{Working in a basis where this is not fulfilled is possible but makes the kinetic term of $\phi_{\rep{r}_1}$ non-diagonal
which one typically wants to avoid.}
In this case it is easy to see that $U_{{\overline{\rep{r}}_1},\rep{r}_1}=U_{\rep{r}_1,{\overline{\rep{r}}_1}}^*$
is a solution to the second consistency condition in~\eqref{eq:consistency_exampleCP}. 
Hence, for CP transformations of complex fields Eq.~\eqref{eq:trafo_general} becomes (we again restore the space-time arguments and ensure that they transform as they should under CP)
\begin{equation}\label{eq:trafo_CP}
\phi(x) = \mat{\phir{1}(x)\\\phi_{\overline{\rep{r}}_1}(x)} \longmapsto \mat{0 & U_{\rep{r}_1,\overline{\rep{r}}_1} \\ U_{\rep{r}_1,\overline{\rep{r}}_1}^* & 0}\mat{\phir{1}(\mathcal{P}x)\\\phi_{\overline{\rep{r}}_1}(\mathcal{P}x)}.
\end{equation}
The fact that the second line of this equation is entirely implied by the first line (for $\phi_{\overline{\rep{r}}_1}=\phir{1}^*$) is why we did not explicitly 
consider the transformation of the complex conjugate representations in our earlier discussion of CP and instead wrote more compactly $\phir{1}\mapsto U_{\rep{r}_1} \phir{1}^*$ where here $U_{\rep{r}_1}\equiv U_{\rep{r}_1,\rep{r}_2}\equiv U_{\rep{r}_1,\overline{\rep{r}}_1}$. 
Nontheless we stress that thinking about the CP transformation in the form of
\eqref{eq:trafo_CP} is tremendously useful, in particular if one wants to consider the explicit group extension of $G$ by the outer automorphism,
and we will come back to this in section~\ref{sec:continuousgroups}. 

\subsection{Generalization to multiple flavors}\label{sec:general-flavor}
As before, also for the case of general automorphism there is the possibility of having multiple identical copies of symmetry representations, 
i.e. multiple ``flavors'' of fields that transform identically under the symmetry $G$.

Again, we will collect the ${\Nf}_i$ flavor-copies of a field in representation $\rep{r}_i$ in the vector $\vec\phi_{\rep{r}_i}:=(\phi_{\rep{r}_i}^{1},..,\phi_{\rep{r}_i}^{{\Nf}_i})$.
Of course, all irreps in a cycle \eqref{eq:out_cycle} have to appear with the same flavor multiplicity implying that ${\Nf}_1={\Nf}_2=\dots={\Nf}_{{\Nr}}\equiv{\Nf}$ must be the same for all appearing irreps.
The general transformation then reads
\begin{equation}\label{eq:general-out}
\phi\equiv\mat{\vec\phi_{\rep{r}_1}\\ \vec\phi_{\rep{r}_2}\\ \vdots\\ \vec\phi_{\rep{r}_{{\Nr}}}} \;{\xmapsto{~u~}}\;\mathcal{U}\,\phi~\equiv~
\mat{\mathbf{0} & \mathcal{U}_{\rep{r}_1,\rep{r}_2} & \mathbf{0} & \mathbf{0} &\dots \\ \mathbf{0} & \mathbf{0} & \mathcal{U}_{\rep{r}_2,\rep{r}_3} & \mathbf{0} & \dots \\
\vdots & \vdots & \vdots & \ddots &   \\
\mathcal{U}_{\rep{r}_{{\Nr}},\rep{r}_1} & \mathbf{0} & \mathbf{0} & \mathbf{0} &\dots
}
\mat{\vec\phi_{\rep{r}_1}\\ \vec\phi_{\rep{r}_2}\\ \vdots\\ \vec\phi_{\rep{r}_{{\Nr}}}}\;.
\end{equation}
The same arguments as in section~\ref{sec:Standard_Form_CP} (see also App.~\ref{sec:Standardform_Derivation}) can be made in order to show 
that the transformation matrices $\mathcal{U}_{\rep{r}_i,\rep{r}_j}$ have to factorize into flavor and symmetry-representation space,
\begin{equation}
 \mathcal{U}_{\rep{r}_i,\rep{r}_j}=\mathcal{A}_{\rep{r}_i,\rep{r}_j}\otimes U_{\rep{r}_i,\rep{r}_j}\;,
\end{equation}
where $\mathcal{A}_{\rep{r}_i,\rep{r}_j}$ again is an $(n_i\times n_i)$-dimensional unitary matrix in flavor space.
Note that despite the fact that we can perform unitary basis changes on the irreps and in flavor space, this does not allow us to completely absorb the 
matrices $\mathcal{A}_{\rep{r}_i,\rep{r}_j}$. This can be seen as follows. We consider again a factorized basis change with 
\begin{equation}
 \vec\phi_{\rep{r}_i}'~=~\mathcal{X}_{\rep{r}_i} \vec\phi_{\rep{r}_i}\;,\qquad\text{with}\qquad \mathcal{X}_{\rep{r}_i}~=~X_{\mathcal{A}_{\rep{r}_i}}\otimes X_{U_{\rep{r}_i}}\;.
\end{equation}
The $\mathbbm{1}_{{\Nf}_i}\otimes X_{U_{\rep{r}_i}}$ part of this basis change can, in special cases, be used to realize the ``$u$-eigenbasis'' \eqref{eq:out-eigenbasis} where all $\mathcal{U}_{\rep{r}_i,\rep{r}_j}$ are absorbed. This crucially relies on the property~\eqref{eq:schur-property}. 
The fact that an analogous property does not hold in flavor space implies that it is, in general, 
not possible to simultaneously absorb all $\mathcal{A}_{\rep{r}_i,\rep{r}_j}$. The general basis change in flavor space reads $X_{\mathcal{A}_{\rep{r}_i}}\otimes \mathbbm{1}_{\mathrm{dim}(\rep{r}_i)}$ and transforms the flavor transformation matrices to
\begin{equation}
\mathcal{A}_{\rep{r}_i,\rep{r}_{i+1}}'~=~X_{\mathcal{A}_{\rep{r}_i}} \mathcal{A}_{\rep{r}_i,\rep{r}_{i+1}} X_{\mathcal{A}_{\rep{r}_{i+1}}}^{\dagger}\,,
\end{equation}
where the indices are understood modulo ${\Nr}$. Hence, the simplest possible form that can be achieved, in general, is given by
\begin{equation}
 \mathcal{A}_{\rep{r}_1,\rep{r}_{2}}'~=~\dots~=~\mathcal{A}_{\rep{r}_{{\Nr}-1},\rep{r}_{{\Nr}}}'~=~\mathbbm{1}_{{\Nf}}\;,\qquad\text{and}\qquad\mathcal{A}_{\rep{r}_{{\Nr}},\rep{r}_{1}}'~=~\Lambda\;,
\end{equation}
where $\Lambda$ is a diagonal matrix with eigenvalues of modulus $1$ defined via
\begin{equation}
\Lambda~:=~X_{\mathcal{A}_{\rep{r}_1}}\,\mathcal{A}_{\rep{r}_1,\rep{r}_{2}}\,\dots\,\mathcal{A}_{\rep{r}_{{\Nr}},\rep{r}_{1}}\,X_{\mathcal{A}_{\rep{r}_1}}^\dagger\;.
\end{equation}
This shows that a possible flavor space standard form of the transformation \eqref{eq:general-out} may be chosen as 
\begin{equation}
\mathcal{U}~=~\mat{\mathbf{0} & \mathbbm{1}_n\otimes U_{\rep{r}_1,\rep{r}_2} & \mathbf{0} & \mathbf{0} &\dots \\ \mathbf{0} & \mathbf{0} & \mathbbm{1}_n\otimes U_{\rep{r}_2,\rep{r}_3} & \mathbf{0} & \dots \\
\vdots & \vdots & \vdots & \ddots &   \\
\Lambda\otimes U_{\rep{r}_n,\rep{r}_1} & \mathbf{0} & \mathbf{0} & \mathbf{0} &\dots
}\;.
\end{equation}
In addition one might also use basis changes of the form $\mathbbm{1}_{{\Nf}_i}\otimes X_{U_{\rep{r}_i}}$ to simplify the form of the $U_{\rep{r}_i,\rep{r}_j}$'s as discussed in the previous section. Other choices of flavor basis may be more practical depending on the application, but this shows that the flavor space rotation can, in general, not be removed. Hence, non-trivial transformations in flavor space may lead to new symmetries in the horizontal space. 
Again we recall that the flavor space transformation matrices $\mathcal{A}_{\rep{r}_i,\rep{r}_j}$ may, in general, not be freely chosen at will but they can be derived from the elementary or composite nature of the states in $\phi_{\rep{r}_i}$.
Note that in the above basis, an $n$-fold application of the total transformation $u$ leads to a transformation $\vec\phi_{\rep{r}_i}\xmapsto{u^n=\mathrm{id}} \left(\Lambda\otimes\mathbbm{1}_{\mathrm{dim}(\rep{r}_i)}\right)\vec\phi_{\rep{r}_i}$.
Hence, the order of the total combined flavor and outer automorphism transformation can straightforwardly be computed as the least common multiple of the order of the automorphism $u$ and the order of the eigenvalues of $\Lambda$.
In the next section we will show an example for a transformation of a composite state, which obeys a non-trivial and non-absorbable flavor space transformation.

\subsection{Example: \texorpdfstring{$\boldsymbol{D_8}$}{D8}}\label{sec:D8}
\begin{table}
\centering
\begin{tabular}{l|rrrrr}
$D_8$ & $\mathrm{id}$ & $(ts)^2=(st)^2$ & $\{ts, st\}$ & $\{s, tst\}$ & $\{t, sts\}$ \\
\hline
size & 1 & 1 & 2 & 2 & 2 \\
order & $1$ & $2$ & $4$ & $2$ & $2$ \\
\hline
$\rep{1}_{++}$ & $1$ & $1$& $1$& $1$& $1$\\
$\rep{1}_{--}$ & $1$ & $1$& $1$& $-1$& $-1$\\
$\rep{1}_{+-}$ & $1$ & $1$& $-1$& $1$& $-1$\\
$\rep{1}_{-+}$ & $1$ & $1$& $-1$& $-1$& $1$\\
$\rep{2}$ & $2$ & $-2$ & $0$ & $0$ & $0$ \\
\hline
\end{tabular}
\caption{\label{tab:chartab-D8}
Character table of $D_8$. In the first line we show the complete conjugacy classes, in the second line their size and in the third line the order of the respective elements.}
\end{table}
To illustrate the treatment of non-CP outer automorphisms of the preceding sections we consider the dihedral group of order $8$, $D_8$, as an example. 
A minimal generating set of $D_8$ is given by\footnote{%
This presentation, which is much favorable for the discussion of the action of the outer automorphism, is related to the more common presentation of $D_8=\Braket{a,b|a^4=b^2=abab=\mathsf{e}}$ via $s=b$, $t=ab$ (see e.g.~\cite[App.A]{Bischer:2020nwl} or \cite{Ishimori:2010au} where the group is called $D_4$).}
\begin{equation}\label{eq:D8Presentation}
D_8~=~\Braket{s,t\,|\,
s^2\,=\,t^2\,=\left(ts\right)^4\,=\,\mathsf{e}}\;.
\end{equation}
$D_8$ has four real one-dimensional and one real two-dimensional irrep as shown in Tab.~\ref{tab:chartab-D8}.
The automorphism structure of $D_8$ can be summarized as
\begin{align}
\mathrm{Centre}(D_8)&~=~\Z2\;,& \mathrm{Aut}(D_8)&~=~D_8\;,&\\
\mathrm{Inn}(D_8)&~=~\Z2\times\Z2\;,&  \mathrm{Out}(D_8)&~=~\Z2\;.&
\end{align}
There is a non-trivial $\Z2$ outer automorphism which we will call $u$ in the following. 
According to the classification of~\cite{Chen:2014tpa}, $D_8$ is a group of type~II~A,
meaning that it has a class-inverting automorphism. Note that it is \textit{not} $u$ that is class-inverting. 
Instead, it is the identity automorphism which is class-inverting, as can easily be seen from the fact that every element of 
$D_8$ is in a conjugacy class with its inverse. This implies that $u$ can give rise to transformation behavior with non-trivial matrices
and we will focus on this outer automorphism in the following.

The simplest representative of $u$ acts as permutation of the two generators $s$ and $t$, 
\begin{equation}
u:\, \left\{s,t\right\}~\xmapsto{~u~}~\left\{t,s\right\}\;.
\end{equation}
Acting on the conjugacy classes, $u$ acts as a permutation of the two right-most conjugacy classes in Tab.~\ref{tab:chartab-D8}, implying that
it should act on the representations as a permutation of $\rep{1}_{+-}\leftrightarrow\rep{1}_{-+}$. In agreement with that, the consistency
conditions 
\begin{align}\label{eq:consistency1pm}
\rho_{\rep{1}_{+-}}(u(g))~&=~U_{\rep{1}_{+-},\rep{1}_{-+}}\,\rho_{\rep{1}_{-+}}(g)\,U^\dagger_{\rep{1}_{+-},\rep{1}_{-+}}\quad \forall g\,,& \\
\rho_{\rep{1}_{-+}}(u(g))~&=~U_{\rep{1}_{-+},\rep{1}_{+-}}\,\rho_{\rep{1}_{+-}}(g)\,U^\dagger_{\rep{1}_{-+},\rep{1}_{+-}}\quad \forall g\,,&
\end{align}
are trivially solved for $U_{\rep{1}_{+-},\rep{1}_{-+}}=U_{\rep{1}_{-+},\rep{1}_{+-}}=1$, while any other attempt at transforming the representations, for example mapping the irreps to themselves, 
has no solution. The consistency conditions for the other one-dimensional irreps read
\begin{align}
    \rho_{\rep{1}_{++}}(u(g))~=~\rho_{\rep{1}_{++}}(g) \,,\qquad  \rho_{\rep{1}_{--}}(u(g))~=~\rho_{\rep{1}_{--}}(g)\,, \qquad \forall g\,,
\end{align}
and are also trivially solved.
Hence we can summarize
\begin{equation}\label{eq:d8-1dus}
U_{\rep{1}_{++}}~=~\e^{\I \alpha_{++}}\,,\quad U_{\rep{1}_{--}}~=~\e^{\I \alpha_{--}}\,,\quad U_{\rep{1}_{+-},\rep{1}_{-+}}~=~\e^{\I \alpha_{+-}}\,,\quad U_{\rep{1}_{-+},\rep{1}_{+-}}~=~\e^{\I \alpha_{-+}}\,,
\end{equation}
where we have explicitly displayed the free phases which are undetermined by the consistency conditions. However, requiring that $u^2$ acts like the identity transformation also on the states, allows us to constrain $\alpha_{++}=\pm\pi$, $\alpha_{--}=\pm\pi$, as well as $\alpha_{+-}=-\alpha_{-+}$.

As a side remark, we note that $u$ cannot be applied as a meaningful operation on a theory that only contains a field $\phi_{\rep{1}_{+-}}$. 
On the other hand, if the theory contains both, a field $\phi_{\rep{1}_{+-}}$ as well as $\phi_{\rep{1}_{-+}}$, then the operation $u$ is well-defined 
as an outer automorphism of the theory. As a simple example consider the Lagrangian mass terms of such a theory,
\begin{equation}
\lagr_m = -\frac{m_{+-}^2}{2}\phi_{\rep{1}_{+-}}\phi_{\rep{1}_{+-}} -\frac{m_{-+}^2}{2}\phi_{\rep{1}_{-+}}\phi_{\rep{1}_{-+}}\,.
\end{equation} 
As $u$ exchanges $\phi_{\rep{1}_{+-}}$ and $\phi_{\rep{1}_{-+}}$, it is clear that $m_{+-}^2\neq m_{-+}^2$ breaks $u$-symmetry explicitly, while leaving out the field
$\phi_{\rep{1}_{-+}}$ would break $u$-symmetry explicitly and maximally.

The two-dimensional representation of $D_8$ can be generated, for instance, by 
\begin{equation}\label{eq:st_generators}
\rho_{\mathbf{2}}(s)~=~\mat{0&1\\1&0},\qquad \rho_{\mathbf{2}}(t)~=~\mat{0&\I\\-\I&0}.
\end{equation}
The transformation under $u$ can be derived from the consistency conditions
\begin{align}
 \rho_{\mathbf{2}}(u(s))~=~\rho_{\mathbf{2}}(t) ~&=~ U_{\mathbf{2}}\,\rho_{\mathbf{2}}(s)\,U_{\mathbf{2}}^\dagger\,,\\
 \rho_{\mathbf{2}}(u(t))~=~\rho_{\mathbf{2}}(s) ~&=~ U_{\mathbf{2}}\,\rho_{\mathbf{2}}(t)\,U_{\mathbf{2}}^\dagger\,,
\end{align}
which are solved by\footnote{%
As discussed in general below Eq.~\eqref{eq:out-eigenbasis}, $U$ cannot be entirely removed by a basis change. The best that can be achieved is diagonalization, in which case
$U'=\mathrm{diag}\left(-1,1\right)$ and in this basis $\rho_{\mathbf{2}}(s/t)'=\left\{\left\{-1,\pm\I\right\},\left\{\mp\I,1\right\}\right\}$. There is 
no particular advantage of working in this basis, which is why we stick to the basis defined by Eq.~\eqref{eq:st_generators}.
}
\begin{equation}\label{eq:d8-2dus}
U_{\mathbf{2}}~=~\mat{0&\eta\\ \eta^{-1}&0}\,\qquad\text{with}\qquad \eta:=\mathrm{e}^{\I\pi/4}\;.
\end{equation}
Again, the global phase of $U_{\mathbf{2}}$, say $\e^{\I\alpha_{\rep{2}}}$, is not fixed from the consistency condition, but it can be fixed to 
$\alpha_{\rep{2}}=\pm\pi$ from the requirement of $U_{\mathbf{2}}^2=\mathbbm{1}$ which would lead to an order-two transformation of the elementary fields in $\rep{2}$ 
in consistency with the order-two outer automorphism $u$.

\paragraph{1. Product representation of $\boldsymbol{D_8}$.}
Let us now give an example for how to investigate the transformation behavior of composite versus elementary states. 
Consider the product state of three $D_8$ doublet states,
\begin{equation}
\mathbf{2}\otimes\mathbf{2}\otimes\mathbf{2}~=~\mathbf{2}\oplus\mathbf{2}\oplus\mathbf{2}\oplus\mathbf{2}\;.  
\end{equation}
The symmetry generators act on this state as $\rho(s)_{\rep{2}}^{\otimes3}$ and $\rho(t)_{\rep{2}}^{\otimes3}$ while the 
outer automorphism acts as $\mathcal{U}_{\rep{2}}=U_{\rep{2}}^{\otimes3}$. The action of these transformations can be simultaneously 
block-diagonalized by a basis change with $\mathcal{X}_{\rep{2}\otimes\rep{2}\otimes\rep{2}}$ that can be found, for example, using the SBD algorithm~\cite{Bischer:2020nwl}
and which is given by the permutation matrix\footnote{%
We will in the following drop the trivial first line in stating permutation matrices.}
\begin{equation}
 \mathcal{X}_{\rep{2}\otimes\rep{2}\otimes\rep{2}}~=~\begin{pmatrix} 1 & 2 & 3 & 4 & 5 & 6 & 7 & 8 \\ 8 & 1 & 3 & 6 & 5 & 4 & 2 & 7 \end{pmatrix}\;.
\end{equation}
In the block-diagonal basis $\mathcal{U}_{\rep{2}}'=\mathcal{X}_{\rep{2}\otimes\rep{2}\otimes\rep{2}}\,\mathcal{U}_{\rep{2}}\,\mathcal{X}^{\dagger}_{\rep{2}\otimes\rep{2}\otimes\rep{2}}$, the transformation of the outer automorphism reads
\begin{equation}
 \mathcal{U}_{\rep{2}}'~=~U_{\rep{2}}\oplus U_{\rep{2}} \oplus U_{\rep{2}}\oplus\left(-\right)U_{\rep{2}}~=~\mathcal{A}_{\rep{2}}\otimes U_{\rep{2}}=\mathrm{diag}\left(1,1,1,-1\right)\otimes U_{\rep{2}}\;,
\end{equation}
where in the last step we have read off the flavor space transformation matrix $\mathcal{A}_{\rep{2}}$. 
Irrespectively of the free choice of phase for the transformation of the elementary doublets we see that there is a 
relative phase of $-1$ that singles out one of the composite doublets. This relative phase
is physical and cannot be removed by a basis change. This example of a different transformation behavior of elementary 
and composite doublets under the transformation of the outer automorphism illustrates 
the possibility to distinguish composite from elementary doublet states.\footnote{%
The group $D_8$ has recently been used in a concrete particle physics model extension of the Higgs sector in order to reduce the fine-tuning of the electroweak scale~\cite{Csaki:2022zbc}.
It would be interesting to investigate the action of the \Z{2} outer automorphism in this model and whether it could be used to detect compositeness.}

\paragraph{2. Regular representation of $\boldsymbol{D_8}$.}
An example for a non-product representation where composite representations transform differently than elementary irreps is given by 
the regular representation of $D_8$ which has a branching
\begin{equation}\label{eq:regD8}
 \mathrm{reg}_{D_8}~=~\mathbf{1}_{++}\oplus\mathbf{1}_{--}\oplus\mathbf{1}_{+-}\oplus\mathbf{1}_{-+}\oplus\rep{2}\oplus\rep{2}\;.
\end{equation}
In the basis $\left\{\mathrm{id},s,ts,sts,\left(ts\right)^2,tst,st,t\right\}$ the regular representation
of the generators $s$ and $t$ as well as the action of the outer automorphism $u$ is given by
\begin{align}
 \rho(s)_\mathrm{reg}~&=~\begin{pmatrix}~2&1&4&3&6&5&8&7~\end{pmatrix}\;,& \\
 \rho(t)_\mathrm{reg}~&=~\begin{pmatrix}~8&3&2&5&4&7&6&1~\end{pmatrix}\;,& \\
 \rho(u)_\mathrm{reg}~&=~\begin{pmatrix}~1&8&7&6&5&4&3&2~\end{pmatrix}\;.& 
\end{align}
The action of the representation matrices is block-diagonalized to realize the decomposition into irreps, Eq.~\eqref{eq:regD8},
by a basis transformation with
\begin{equation}
 \mathcal{X}_\mathrm{reg}~=~
 \frac{1}{2\sqrt{2}}
 \left(
\begin{array}{rrrrrrrr}
 -1 & -1 & -1 & -1 & -1 & -1 & -1 & -1 \\
 -1 & 1 & -1 & 1 & -1 & 1 & -1 & 1 \\
 1 & 1 & -1 & -1 & 1 & 1 & -1 & -1 \\
 1 & -1 & -1 & 1 & 1 & -1 & -1 & 1 \\
 \eta^{-1} & -\I & \eta & -1 & -\eta & \I & -\eta^{-1} & 1 \\
 -\I & \eta^{-1} & -1 & \eta & \I & -\eta & 1 & -\eta^{-1} \\
 -\eta & -\I & -\eta^{-1} & -1 & \eta^{-1} & \I & \eta & 1 \\
 -\I & -\eta & -1 & -\eta^{-1} & \I & \eta^{-1} & 1 & \eta \\
\end{array}
\right)\;.
\end{equation}
In the block-diagonal basis, the transformation matrix of the outer automorphism is given by
\begin{equation}
 \mathcal{U}_{\mathrm{reg}}~=~1\oplus 1 \oplus \begin{pmatrix}0 & 1 \\ 1& 0\end{pmatrix}
 \oplus U_{\rep{2}} \oplus -U_{\rep{2}}\;.
\end{equation}
The transformation of the singlets $\rep{1}_{\pm\mp}$ is consistent with the expectation from Eq.~\eqref{eq:consistency1pm}.
For the doublets, we can again read off the flavor space transformation matrix 
\begin{equation}
 \mathcal{U}_{\rep{2}}~=~\mathcal{A}_{\rep{2}}\otimes U_{\rep{2}}~=~\mathrm{diag}\left(1,-1\right)\otimes U_{\rep{2}}\;.
\end{equation}
This shows that also the regular representation contains a composite doublet state that transform with an irremovable relative phase 
with respect to an elementary doublet state.

\section{Semisimple compact Lie groups}
\label{sec:continuousgroups}
Finally, let us extend our discussion also to a particularly interesting special case of continuous groups, namely the compact semisimple groups.
In principle, general CP transformations of those groups are of the form $\phi(x)\mapsto \mathcal{U}\phi^*(\mathcal{P}x)$ and have already been exhaustively discussed in~\cite{Grimus:1995zi}.\footnote{%
While~\cite{Arias-Tamargo:2019jyh,Henning:2021ctv} (see also~\cite{Bourget:2018ond}) refer to the complex conjugation outer automorphism of $\SU{N}$ as charge conjugation $\mathcal{C}$
we refer to the very same outer automorphism as CP. At the level of the transformation of $\SU{N}$ the difference in calling one and the same outer automorphism 
charge conjugation $\mathcal{C}$ vs.\ a CP transformation is purely semantics. However, we note that a strict transformation $\phi_{\rep{r}_i}\mapsto U\phi_{\rep{r}_i}^*$ for all fields, in general, corresponds to a complex conjugation outer automorphism also of the Lorentz group representations, hence, to a CP transformation, see e.g.~\cite{Buchbinder:2000cq}. 
The fact that the actual $\mathcal{C}$ transformation does not correspond to a pure complex conjugation map but requires additional 
non-trivial transformations between fields can simply be seen from the fact that, unlike CP, charge conjugation can be broken maximally by simply leaving out representations, as is the case in the SM.}
However, there have been claims in the recent literature~\cite{Arias-Tamargo:2019jyh,Henning:2021ctv} that for groups of the type $\SU{2N}$ there would be a distinct set of outer automorphism transformations possible, 
which transforms irreducible representations $\phi_{\rep{r}}\mapsto U\phi_{\rep{r}}^*$ with antisymmetric matrices $U^{\mathrm{T}}=-U$.
At a first look this seems to contradict the general statement of~\cite{Grimus:1995zi}, that a CP basis with $U=\mathbbm{1}$ can always be found.\footnote{%
We recall that under a basis change, $U$ transforms with a UCT, see section~\ref{sec:Standard_Form_CP}, such that the transposition behavior of $U$ is preserved.}
Since a transformation behavior with non-trivial $U$ would be highly interesting, also from the point of view of the transformation of composite states, we 
carefully investigate the case here in order to clarify the recent claims. 
The general conclusion is that the antisymmetric outer automorphism transformation matrices for $\SU{2N}$ are given by specific choices of \textit{inner} automorphisms in addition to the unique outer automorphism. Consequently, transformations with these antisymmetric matrices do not correspond to new
outer automorphisms. In turn, this implies that product composite states of $\SU{N}$ will always transform under outer automorphisms of $\SU{N}$ like their elementary counterparts. We will be as brief as possible here, focusing our attention on the transformation of composite states, while postponing the more general discussion to the forthcoming~\cite{DT:2023}.
For the mathematical details of the discussion about outer automorphisms of semisimple Lie algebras we follow~\cite{Fuchs:1997jv}, 
while a concise introduction with a few explicit examples for outer automorphisms can be found in~\cite[Sec.~3.4]{Trautner:2016ezn}.

A general element of $\SU{N}$ can be written as $M=\e^{\I\theta_a T_a}$ with real constants $\theta_a$ and traceless generators $T_{a}$ where $a=1,\dots,n_{\mathfrak{g}}$.
The generators obey the Lie algebra~$\mathfrak{g}$ 
\begin{equation}\label{eq:Liealgebra}
 \left[T_a,T_b\right]~=~\I\,f_{abc}\,T_c\;,
\end{equation}
with structure constants $f_{abc}$. A general automorphism acts on the generators as 
\begin{equation}
 u:~T_a~\mapsto~R_{ab}\,T_b\;,
\end{equation}
where $R_{ab}$ has to fulfill
\begin{equation}\label{eq:adjoint}
 R_{aa'}\,R_{bb'}\,f_{a'b'c}~=~f_{abc'}\,R_{c'c}\;.
\end{equation}
If there is a representation generated by $\left\{T_a\right\}$, one can show that also $\left\{-T^\mathrm{T}_a\right\}$
fulfill the Lie algebra~\eqref{eq:Liealgebra}, hence, generate a representation of the same dimension as $\left\{T_a\right\}$. 
In a basis with hermitean generators, it is straightforward to see that this is the 
complex conjugate representation (hence, it holds for all other bases). An (outer) automorphism $u$ that maps a representation to its complex conjugate has to fulfill the
consistency condition 
\begin{align}\label{eq:consistency_condition_CG}
    U\,\left(-T_a^{\mathrm{T}}\right)\,U^{\dagger}~=~R_{ab}\,T_b\;,\qquad\forall a\;.
\end{align}
Given a specific representation, the solution for $U$ gives the explicit transformation matrix $\phi_{\rep{r}}\mapsto U\phi_{\rep{r}}^*$.
Note how both, Eq.~\eqref{eq:consistency_condition_CG}, as well as~\eqref{eq:adjoint} for the adjoint representation, are nothing but the 
the general consistency condition Eq.~\eqref{eq:consistency_irrep} applied to this specific case. 
The transformation within the adjoint representation is equivalent to the transformation
of the elements of the abstract algebra, i.e.\ the transformation of the Lie lattice. 
Hence, it is straightforward to distinguish inner from outer automorphisms by $\mathrm{det}(R)=\pm1$,
where the negative sign applies to outer automorphisms. 

Assuming an automorphism $u$ of order $2$ we have $R^2=\mathbbm{1}$ and the consistency condition~\eqref{eq:consistency_condition_CG} can be applied twice to show that
\begin{equation}
 \left[U\,U^*,T_a\right]~=~0\;.
\end{equation}
By Schur's lemma and the unitarity of $U$ this implies that $U^{\mathrm{T}}=\pm U$, i.e.\ $U$ is either symmetric or antisymmetric. 
In any case, one can use a unitary congruence transformation to transform $U$ into a basis where it obeys one of the standard forms discussed 
in section~\ref{sec:Standard_Form_CP}. 

Considering a field $\phi$ and its complex conjugate that transforms under $SU(N)$ as 
\begin{equation}
\begin{pmatrix}\phi \\ \phi^*\end{pmatrix}~\xmapsto{~\SU{N}~}~
\begin{pmatrix} M & 0 \\ 0 & M^* \end{pmatrix}
\begin{pmatrix}\phi\\\phi^*\end{pmatrix}\,,
\end{equation}
then the pair will transform under the involutory complex conjugation outer automorphism $u$ as\footnote{%
We stress that transformations
\begin{equation}
\begin{pmatrix}\phi \\ \phi^*\end{pmatrix}\mapsto
\begin{pmatrix} 0 & U \\ U^{-1} & 0 \end{pmatrix}
\begin{pmatrix}\phi\\\phi^*\end{pmatrix}\;,
\end{equation}
as advertised in~\cite{Arias-Tamargo:2019jyh,Henning:2021ctv}, are generally \textit{not} self-consistent. 
They are self-consistent if and only if $U^{-1}=U^*$, i.e.\ transformations of order $2$ and unitary \textit{symmetric} matrices.
This view was very recently confirmed in~\cite{Kondo:2022wcw} which appeared while our paper was in the final stages of preparation,
see \textit{note added} at the end of this work.}
\begin{equation}\label{eq:SUNout}
\begin{pmatrix}\phi \\ \phi^*\end{pmatrix}~\xmapsto{~u~}~
\begin{pmatrix} 0 & U \\ U^*& 0 \end{pmatrix}
\begin{pmatrix}\phi\\\phi^*\end{pmatrix}\;.
\end{equation}
Whether or not $U$ in this transformation is symmetric or antisymmetric crucially depends on the choice of 
automorphism $u$, or in more detail, the choice of inner automorphism to accompany the outer automorphism $u$.\footnote{%
We emphasize that strictly speaking an outer automorphism is a coset of automorphisms. 
Hence, despite the fact that for practical purpose we always fix an associated inner automorphism
the correct way of thinking is that each outer automorphisms contains \textit{all} inner automorphisms.
}
For the groups $\SU{2N+1}$ it happens that there is no option to chose an inner automorphism with antisymmetric $U$ (this is prohibited simply by the fact
that there are no full ranked antisymmetric matrices of odd dimension). However, for the groups of type $\SU{2N}$ it is possible. 
A simple example is the $\SU{2N}$ group element
\begin{equation}\label{eq:ASgroupelement}
\Sigma~:=~\Sigma_-\oplus\dots\oplus\Sigma_-\;, 
\end{equation}
with $\Sigma_-$ defined in Eq.~\eqref{eq:sigma-}. We will refer to the corresponding inner automorphism as $u_\Sigma$ and its explicit action 
on the adjoint representation as $R_\Sigma$ (the consistency condition for $u_\Sigma$ as an inner automorphism reads 
$\Sigma\,T_a\,\Sigma^\dagger=(R_\Sigma)_{ab}T_b$). $u_\Sigma^2=1$ is involutory corresponding to $R_{\Sigma}^2=\mathbbm{1}$ (this is not in contradiction to 
the fact that the action of $u^2_\Sigma$ on fields in the fundamental representation, given by $\Sigma^2=-\mathbbm{1}$, only closes to an element in the center of the group). 

Assuming that $U$ would be symmetric, and we had already chosen a basis such that $U=\mathbbm{1}$ (which is always possible, see the discussion of standard forms in section~\ref{sec:Standard_Form_CP}) it is clear that a composition of transformations $\widetilde{u}:=u_\Sigma\cdot u$ will lead to an outer automorphism $\widetilde{u}$ that transforms $\phi\mapsto \widetilde U\phi^*$ with an antisymmetric matrix $\widetilde{U}:=\Sigma U$.

Likewise, if we start with an antisymmetric $U$, we can choose a basis such that $U$ obeys the standard form discussed in section~\ref{sec:Standard_Form_CP},
implying that it is given by the same matrix as stated in Eq.~\eqref{eq:ASgroupelement}. In this case, a composition of the outer automorphism with the inner automorphism $u_\Sigma$
would remove the antisymmetric transformation matrix and lead to an outer automorphism transforming the states with the identity matrix.

Hence, the difference between symmetric and antisymmetric transformation behavior
for outer automorphisms of groups of the type $\SU{2N}$ is an inner automorphism. Consequently, the groups generated by 
\begin{equation}\label{eq:GammaU}
 \Gamma_U~=~\left\{
 \begin{pmatrix} M & 0 \\ 0 & M^* \end{pmatrix},
 \begin{pmatrix} 0 & U \\ U^* & 0 \end{pmatrix} 
 \Bigg| 
 M\in\SU{N}
 \right\}\;,
\end{equation}
or 
\begin{equation}\label{eq:GammaUtilde}
 \Gamma_{\widetilde{U}}~=~\left\{
 \begin{pmatrix} M & 0 \\ 0 & M^* \end{pmatrix},
 \begin{pmatrix} 0 & \widetilde{U} \\ \widetilde{U}^* & 0 \end{pmatrix} 
 \Bigg| 
 M\in\SU{N}
 \right\}\;,
\end{equation}
are exactly the same group.\footnote{%
The second element in \eqref{eq:GammaU} can be generated from the second element in \eqref{eq:GammaUtilde} by using the group
element $M^{-1}\oplus M^{\mathrm{T}}$. The second element in \eqref{eq:GammaUtilde} can be generated from the second element in \eqref{eq:GammaU}
by using the group element $M\oplus M^*$.}

One may even go further and consider inner automorphisms of higher order. In this 
case the representation matrix $\widetilde U=MU$ may not have definite symmetry properties at all. 
Nontheless, for \SU{N} it is guaranteed that every outer automorphism squares to an 
inner one, $u^2=\mathrm{inn}$. The easiest way to see this, is by noticing that the 
according transformation matrix $\phi\xmapsto{u^2}(\widetilde{U}\widetilde{U}^*)\phi$ is unitary 
and, therefore, a group element of \SU{N}.\footnote{%
This situation can be different for other groups. For example, for finite groups of the type~II~B even outer automorphisms that square
to inner automorphisms can lead to a non-trivial (in this case Abelian) enhancement of the symmetry group~\cite{Chen:2014tpa}.}

For completeness, we remark why these statements are in no contradiction with Grimus and Rebelo~(GR)~\cite{Grimus:1995zi},
who claimed that for groups of $\SU{N}$ one can always find a CP basis with $U=\mathbbm{1}$. 
For this, one has to note that GR define CP transformations via a special choice of outer automorphism
that automatically fixes the admissible choices of inner automorphisms. The automorphism chosen by GR is the so-called 
contragredient automorphism  $u_\Delta$ (also Weyl automorphism or Chevally involution~\cite{Fuchs:1997jv}) that acts as 
an inversion (point reflection on origin) of the root lattice of $\mathfrak{g}$. This is motivated 
by the fact that $u_\Delta$ leads to the reflection, or inversion, of ``all'' quantum numbers.
In this case one can show, see~\cite[App.~F, Case 2]{Grimus:1995zi}, that only the symmetric case $U^{\mathrm{T}}=U$
is allowed, implying that one can always find a so-called CP basis where $U=\mathbbm{1}$. 
Note that while the requirement of inversion of ``all'' quantum numbers (QNs) is sufficient to find a physical CP transformation
it is not necessary, see the extensive discussion in~\cite[Sec.~VII.13.2]{Fallbacher:2015upf}. 
In particular, also the contragredient automorphism does not inverse \textit{all} QNs. For example, 
it does not inverse the eigenvalues of the quadratic or any other even-order Casimir operator.
At the same time, applying additional inner automorphisms (i.e.\ symmetry transformations), such as for example $u_\Sigma$, 
must not change physical predictions. Even though a composition $u_\Sigma\cdot u_\Delta$ does not correspond to a root lattice inversion
(i.e.\ it does not invert ``all'' quantum numbers), it still does map the quantum numbers of a state in an irrep $\rep{r}$ to the 
quantum numbers of a state in the complex conjugate irrep $\rep{r}^*$.
The conclusion is that allowing only the contragredient automorphism as physical CP transformation is too strict. A sufficient condition 
for a physical CP transformation is the use of a class-inverting automorphism~\cite{Chen:2014tpa, Fallbacher:2015upf, Trautner:2016ezn}.

This whole discussion does not affect the possibility that different subgroups of $\SU{N}$ might be anomalous~\cite{Henning:2021ctv}. 
Such anomalies might single out one or the other inner automorphism above others. However, the fact that one or the other symmetry element
can be broken (softly or explicitly by the anomaly) may always be the case, but this discussion is somewhat orthogonal to the discussion about outer automorphisms and so we do not pursue it further.

Finally, we are coming back to the distinction of composite versus elementary states under outer automorphisms. 
Our discussion in this section shows that, up to a choice of inner automorphism, 
the outer automorphism transformation matrices of $\SU{N}$ irreps may always be brought to a form $U_{\rep{r}_i}=\mathbbm{1}$.
In particular, the sole $\Z2$ outer automorphism of \SU{N} ($N\neq2$) is class-inverting and involutory.
In this sense $\SU{N}$ groups closely resemble the finite groups of type~II~A~\cite{Chen:2014tpa}.
The outer automorphism transformation matrix of a composite product state, $\mathcal{U}$, can always derived from the 
direct product of transformation matrices of elementary states, $\mathcal{U}=U_{\rep{r}_i}\otimes U_{\rep{r}_j}\otimes\dots$,
and such a product preserves the transformation behavior under transposition. This shows that, unfortunately, no 
non-trivial transformation of composite product states is possible for representation of $\SU{N}$.
If one would, by accident or maliciously, choose a ``wrong'' inner automorphism to accompany the outer one,
one may find non-trivial (e.g.\ antisymmetric) transformation behavior of composite states. However, it is immediately clear that such transformations
must correspond to inner automorphisms, i.e.\ symmetry transformations, which cannot distinguish between composite and 
elementary states. Of course, this does not preclude the omnipresent possibility of \textit{imposing} non-trivial transformation behavior 
of fields with matrices $\mathcal{A}$ in flavor space, once identical copies of elementary states are introduced.

As an important caveat, note that our statements do not exclude non-trivially transforming \textit{non-product} composite states. Those could transform with anti- or mixed-symmetric matrices $\mathcal{U}$
which would, together with the always symmetric transformation behavior of contained irreps, 
immediately signal a necessarily non-trivial transformation behavior in flavor space. However, this discussion and treatment of explicit examples is beyond the scope of this work, 
as we are presently not aware of any construction of non-product composite states of \SU{N} (or exclusion thereof).

\section{Conclusions}
\label{sec:conclusions}
We have discussed composite product and non-product representations of groups and their transformation behavior under
outer automorphisms of the groups (such as P, C, CP but also outer automorphisms in general). 
It is a common occurrence that composite states and elementary states which transform in exactly the same way under group transformations, 
can be distinguished by their different transformation behavior under outer automorphisms of the group. 
This different transformation behavior under outer automorphisms can serve to distinguish composite from elementary states at a long distance, 
i.e.\ without performing a short range scattering experiment. 

We have clearly identified situations in which composite states transform differently than their elementary counterparts. Nontheless, we note that a positive proof of compositeness, in general, depends on an assumption about the transformation behavior permitted for elementary states. That is, only if the absence of certain elementary states is assumed, it is possible to lead a positive proof of compositeness. This is already true in well-known standard cases such as, for example, quark anti-quark bound state mesons. This does not negatively effect our discussion because we focus on the \textit{relative} difference in transformation behavior between elementary and composite states.
On the other hand, we have pointed out that the hypothesis of compositeness can be unequivocally rejected in settings where all composite states transform non-trivially 
compared to an elementary state in the same group representation. 

It was central to our discussion to establish a standard form of general CP transformations (and outer automorphisms in general) for symmetry representation and flavor spaces. 
We have given easy to comprehend examples for several types of non-trivial CP and non-CP outer automorphisms for product and non-product states based on the groups $\Sigma(72)$ and $D_8$. 
While these are finite groups, we stress that our arguments are general and analogous discussion should be done to further investigate infinite and continuous groups.

This is exemplified by the discussion in section~\ref{sec:continuousgroups}, where we have critically investigated a recent claim that new outer automorphisms with 
antisymmetric transformation matrices would exist for $\SU{2N}$ groups. We have shown that those transformations are always related by an inner automorphism to the standard, order-two outer automorphism of \SU{N} that transforms representations with symmetric matrices. Hence, no new non-trivial CP transformations are possible for $\SU{2N}$. With the discussion of the present paper, this also excludes the possibility of having non-trivially 
transforming composite product states for $\SU{N}$ groups. This does not preclude the existence of non-trivially transforming \textit{non-product} composite states of $\SU{N}$, and this is an interesting topic for future research.

The somewhat general discussion of this paper is expected to have important applications in many directions. For example, for classifications of exotic four, five and higher-multiplicity direct product quark bound states, their transformations under flavor symmetries and the corresponding outer automorphisms. 
Also interesting will be the application of our methods to the outer automorphism transformation behavior of non-product composite states of the Poincar\'e group.
Since non-product composite states are entangled states, we also envisage new directions in the treatment of outer automorphism of symmetries of general quantum states in basic quantum mechanics and applications to quantum information theory. Ultimately, our insights will also help to scrutinize the compositeness of what we currently believe to be the elementary particles of Nature. This would be tremendously useful, in particular, if compositeness and the respective short distance physics would arise at scales smaller than what we can hope to probe at colliders in the foreseeable future.

 \section*{Acknowledgements}
The authors would like to thank Michael~A.~Schmidt for his support with the \textsc{Discrete}-package \cite{Holthausen:2011vd} of which we made extensive use.
AT would like to thank Michael Ratz for discussions that led to the initial idea for the present paper, as well as Ofri Telem for directing our attention to electric-magnetic dressed states.
IB acknowledges support by the IMPRS for Precision Tests of Fundamental Symmetries. The work of CD is supported by the F.R.S./FNRS under the Excellence of Science (EoS) project No.\ 30820817 - be.h ``The H boson gateway to
physics beyond the Standard Model'' and by the IISN convention No.\ 4.4503.15.
The work of AT was in part performed at the Aspen Center for Physics, which is supported by National Science Foundation grant PHY-1607611.
The work of AT in Aspen was partially supported by a grant from the Simons Foundation.

 \section*{Note added}
During the completion of our work, Ref.~\cite{Kondo:2022wcw} appeared on the arXiv. 
Ref.~\cite{Kondo:2022wcw} corrects the inconsistent form of the representation matrix of the 
$\SU{N}$ outer automorphism of~\cite{Arias-Tamargo:2019jyh,Henning:2021ctv}
and, therefore, agrees with the well-known general form of the representation matrix given in~\eqref{eq:SUNout}.
Nonetheless, we still disagree with several statements in~\cite{Kondo:2022wcw}~(and~\cite{Arias-Tamargo:2019jyh,Henning:2021ctv}). Most importantly,
we have shown explicitly in section~\ref{sec:continuousgroups} that outer automorphisms of $\SU{2N}$
with symmetric and antisymmetric transformation matrices do \textit{not} lead to different extensions of $\SU{2N}$.
This is the case, simply because the symmetric and antisymmetric transformations only differ by an inner automorphism 
(which acts on the outer automorphism matrix by multiplication, \textit{not} conjugation).
Since there seems to be a lot of confusion in the recent literature about these, in principle, 
settled topics (see~\cite{Grimus:1995zi} and other references of the present paper), 
we will give a more extensive treatment in the forthcoming paper~\cite{DT:2023}.

\appendix
\section*{Appendix}
\section{\boldmath Motivation for the factorization of \texorpdfstring{$\mathcal{U}_{\rep{r}_i}$}{Uri}}
\label{sec:Standardform_Derivation}
Here we give additional details on the factorization of a matrix $\mathcal{U}_{\rep{r}_i}$ in Eq.~\eqref{eq:acal}.
Let us write the matrix in the form
\begin{equation}
\mathcal{U}_{\rep{r}_i}=\mat{A_{11} & \dots & A_{1n_i}\\\vdots & \ddots&\vdots\\A_{n_i1}&\dots&A_{n_in_i}},
\end{equation}
where $A_{\alpha\beta}$ are $\left(\dim({\rep{r}_i})\times\dim({\rep{r}_i})\right)$-dimensional matrices. 
Then the consistency condition Eq.~\eqref{eq:consistency} requires that
\begin{equation}\label{eq:wconsistency}
\rho_{\rep{r}_i}^{\oplus n_i}\!\left(u(g)\right)\;\mathcal{U}_{\rep{r}_i}~=~\mathcal{U}_{\rep{r}_i}\;\left(\rho_{\rep{r}_i}^{\oplus n_i}(g)\right)^*.
\end{equation}
Here, the block-diagonal matrix $\rho_{\rep{r}_i}^{\oplus n_i}$ is the restriction of $\rho$ acting on the vector $(\phi_{\rep{r}_i}^1,\dots,\phi_{\rep{r}_i}^{n_i})^\mathrm{T}$.
Working out the matrix multiplication, \eqref{eq:wconsistency} can be written as
\begin{equation}
\mat{\rho_{\rep{r}_i}(u(g))A_{11} & ... & \rho_{\rep{r}_i}(u(g))A_{1n_i}\\
\vdots&\ddots&\vdots\\
\rho_{\rep{r}_i}(u(g))A_{n_i1} & ... & \rho_{\rep{r}_i}(u(g))A_{n_in_i}}
=\mat{A_{11}\rho_{\rep{r}_i}^*(g) & ... & A_{1n_i}\rho_{\rep{r}_i}^*(g) \\
\vdots&\ddots&\vdots\\
A_{n_i1}\rho_{\rep{r}_i}^*(g) & ... & A_{n_in_i}\rho_{\rep{r}_i}^*(g)}.
\end{equation}
By comparison of entries, each of the blocks $A_{\alpha\beta}$ must satisfy 
\begin{equation}\label{eq:consistency_A}
\rho_{\rep{r}_i}(u(g))\,A_{\alpha\beta} = A_{\alpha\beta}\,\rho_{\rep{r}_i}^*(g)\;.
\end{equation}
This equation can be rewritten by substituting the consistency condition \eqref{eq:consistency_irrep} on the left-hand side, which yields
\begin{equation}
\rho_{\rep{r}_i}^*(g)\,U_{\rep{r}_i}^\dagger\,A_{\alpha\beta} = U_{\rep{r}_i}^\dagger\,A_{\alpha\beta}\,\rho_{\rep{r}_i}^*(g)\,.
\end{equation}
Schur's lemma then implies that $U_{\rep{r}_i}^\dagger A_{\alpha\beta}$ must be proportional to the identity matrix. Therefore, using the unitarity of $U_{\rep{r}_i}$ we can write
\begin{equation}
A_{\alpha\beta}=a_{\alpha\beta}U_{\rep{r}_i} \qquad \forall\alpha,\beta\,,
\end{equation}
where $a_{\alpha\beta}\in \mathbb{C}$. This shows the factorization $\mathcal{U}_{\rep{r}_i} = \mathcal{A}_{\rep{r}_i} \otimes U_{\rep{r}_i}$ of Eq.~\eqref{eq:acal}.

\section{Standard form of a unitary matrix under unitary congruence transformations}
\label{sec:standard-appendix}
Let $\mathcal{A}$ be a unitary $n\times n$ matrix. Ecker et al.~\cite{Ecker:1987qp} defined a standard form 
of $\mathcal{A}$ that can be reached by transformations with unitary matrices $X$,
\begin{equation}\label{eq:std_form_EGN}
X\mathcal{A}X^\mathrm{T} = \mathbbm{1}_m \oplus \left(\bigoplus_i^{\ell} \mat{\cos \alpha_i & \sin \alpha_i\\
-\sin\alpha_i & \cos \alpha_i}\right),\quad 2\ell+m=n\,.
\end{equation}
Weinberg~\cite[Ch.2, App.C]{Weinberg:1995mt2} uses the equivalent standard form
\begin{equation}\label{eq:std_form_Weinberg}
X\mathcal{A}X^\mathrm{T} = \mathbbm{1}_m \oplus \left(\bigoplus_i^{\ell} \mat{0&\mathrm{e}^{\I\alpha_i}\\\mathrm{e}^{-\I\alpha_i} &0 }\right),\quad 2\ell+m=n\,.
\end{equation}
In the original proof of~\cite{Ecker:1987qp}, $m$, $\ell$, and $\alpha_i$ are determined by the eigenvalues of the matrix
\begin{equation}
\frac{1}{4}(\mathcal{A}+\mathcal{A}^\mathrm{T})^\dagger (\mathcal{A}+\mathcal{A}^\mathrm{T}),
\end{equation}
which has an $m$-fold degenerate eigenvalue $1$, and the two-fold degenerate eigenvalues $\cos^2\alpha_i$. 
In case $\cos\alpha_i=0$ the respective block reduces to $\Sigma_-$ defined in \eqref{eq:sigma-}.

Already in~\cite{Ecker:1987qp} it was noted that another simple way of determining $m$ and $\alpha_i$ 
is by determining the eigenvalues of $\mathcal{A}^*\mathcal{A}$. Then $m$ corresponds to the degeneracy of the eigenvalue $1$, while the angles $\alpha_i$ are 
given as half the arguments of the $\ell$ pairwise appearing eigenvalues $\mathrm{e}^{\pm2\I\alpha_{i}}$.

As it might be of wider interest, here we give a different proof for the standard forms
\eqref{eq:std_form_EGN} and \eqref{eq:std_form_Weinberg}, 
based on a general formula for normal matrices given by Youla~\cite{youla_1961}.
A normal matrix is one for which $M^\dagger M=M M^\dagger$.
For a unitary matrix $\mathcal{A}$, the combination $\mathcal{A}^*\mathcal{A}$ is obviously normal.
According to Youla~\cite[Theorem 2]{youla_1961}, then there exists a unitary $X$, such that 
\begin{equation}\label{eq:youla}
X\mathcal{A}X^\mathrm{T} = \Delta \oplus \Omega\,,
\end{equation}
where $\Delta$ is an $m\times m$ upper triangular matrix, and $\Omega$ is a $2\ell\times 2\ell$ block upper triangular matrix 
with blocks on the diagonal being $2\times2$ matrices of the form~\cite{youla_1961}
\begin{equation}
\Sigma_i=\mat{0&\lambda_i/\eta_i \\ \lambda_i^*\eta_i &0}, \quad \eta_i>0\;.
\end{equation}
Furthermore, $\Delta$ and $\Omega$ satisfy
\begin{equation}
\Delta^*\Delta=\mathrm{ diagonal }\;,\qquad\text{and}\qquad 
\Omega^*\Omega=\mathrm{ diagonal}\;.
\end{equation}
From the unitarity of $\mathcal{A}$ it is clear that $\Delta$ and $\Omega$ also must be unitary. Since they are upper triangular (upper block triangular) already, 
they must be diagonal and block-diagonal, respectively. Hence, $\Delta$, corresponds to the symmetric part of $\mathcal{A}$, while $\Omega$ corresponds to the anti- or mixed-symmetric parts.

Possible phases on the diagonal of $\Delta=\mathrm{diag}(\mathrm{e}^{\I\delta_1},\dots,\mathrm{e}^{\I\delta_m})$ can be absorbed by a unitary matrix $Y=\mathrm{diag}(\mathrm{e}^{-\I\delta_1/2},\dots,\mathrm{e}^{-\I\delta_m/2})$ such that $Y\,\Delta\,Y^\mathrm{T}=\mathbbm{1}_m$.
Including $Y$ in the basis change $X$ establishes the symmetric part of the standard form.

In our case, the blocks on the diagonal of $\Omega$ fulfill
\begin{equation}
\mathbbm{1}=\Sigma_i^\dagger\Sigma_i 
= \mat{|\lambda_i|^2\eta_i^2 & 0 \\ 0 & |\lambda_i|^2/\eta_i^2} \quad\Rightarrow\quad 
\eta_i=|\lambda_i|=1\,.
\end{equation}
This shows that $\Sigma_i$ can be written as
\begin{equation}
\Sigma_i = \mat{0&\mathrm{e}^{\I\alpha_i}\\\mathrm{e}^{-\I\alpha_i} &0 }\;.
\end{equation}
The variable name $\alpha_i$ is chosen deliberately, since by rotation with the unitary matrix
\begin{equation}
Z=\sqrt{\frac{-\I}{2}}
\mat{1&\I\\\I&1},
\end{equation}
we recover
\begin{equation}
Z\Sigma_i Z^\mathrm{T} = \mat{\cos \alpha_i & \sin \alpha_i\\
-\sin\alpha_i & \cos \alpha_i}.
\end{equation}
This establishes the antisymmetric and mixed-symmetric parts of both standard forms.

Finally, we clarify the relation of $\alpha_i$ to the eigenvalues of $\mathcal{A}^*\mathcal{A}$.
Note that $\mathcal{A}^*\mathcal{A}$ and $\left(X\mathcal{A}X^\mathrm{T}\right)^* X\mathcal{A}X^\mathrm{T}$
are related by a similarity transformation,
\begin{equation} 
X^*\mathcal{A}^*\mathcal{A}X^\mathrm{T}=
\left(X\mathcal{A}X^\mathrm{T}\right)^* X\mathcal{A}X^\mathrm{T}\;.
\end{equation}
Hence, they have the same eigenvalues. Using the decomposition given in Eq.~\eqref{eq:youla}, $\left(X\mathcal{A}X^\mathrm{T}\right)^* X\mathcal{A}X^\mathrm{T}$ 
is given by $\Delta^*\Delta\oplus\Omega^*\Omega$ and, hence, diagonal. The eigenvalues of $\mathcal{A}^*\mathcal{A}$ are, therefore, the ones of $\Delta^*\Delta\oplus\Omega^*\Omega$
and can simply be read off.
They correspond to the $m$-fold degenerate eigenvalue $1$, as well as the diagonal entries of $\Omega^*\Omega$,
\begin{equation}
\Sigma_i^*\Sigma_i = \mat{\lambda_i^2&0\\0&(\lambda_i^*)^2}\;.
\end{equation}
These come in pairs and correspond to $\e^{\pm 2\I \alpha_i}$.

\section{Details of the type~II~B group \texorpdfstring{$\Sigma(72)$}{Sigma(72)}}\label{app:Sig72SG144}\label{app:Sig72}
Here we give necessary details for the group $\Sigma(72)$ with GAP \texttt{SmallGroupID} $\SG{72,41}$. 
More information on this group can be found in \cite{Damhus:1981, Chen:2014tpa, Trautner:2016ezn}, 
but be aware that we use a different basis in the present work.
A minimal generating set contains two elements, which we call $M=f_1*f_4$ and $P=f_2$, where the $f_i$ denote the generators in the 
standard implementation in GAP~\cite{GAP4}.
The irreps are denoted by $\mathbf{1}_{i=0,1,2,3}$, $\mathbf{2}$, as well as $\mathbf{8}$, and they have the obvious dimensions.
The $\mathbf{2}$ is a pseudo-real representation, while all other irreps are real.
The generators of the one-dimensional representations are $\rho_{\mathbf{1}_i}(M)=1,-1,-1,1$, and $\rho_{\mathbf{1}_i}(P)=1,-1,1,-1$,
while the generators of the two-dimensional representation in our basis read
\begin{equation}\label{eq:S72gens2D}
\rho_{\textbf{2}}(M)=
\begin{pmatrix}
0 & 1\\
-1 & 0\\
\end{pmatrix},\quad\text{and}\quad
\rho_{\textbf{2}}(P)=
\begin{pmatrix}
-\mathrm{i} & 0\\
0 & \mathrm{i}\\
\end{pmatrix}.
\end{equation}
The generators of the octet representation in our basis read
\begin{align}\label{eq:S72gens8D}
\rho_{\mathbf{8}}(M) ~=&~ \frac{1}{2}\,\left(
\begin{array}{cccccccc}
 0 & 0 & 2 & 0 & 0 & 0 & 0 & 0 \\
 0 & 0 & 0 & 2 & 0 & 0 & 0 & 0 \\
 -1 & \sqrt{3} & 0 & 0 & 0 & 0 & 0 & 0 \\
 \sqrt{3} & 1 & 0 & 0 & 0 & 0 & 0 & 0 \\
 0 & 0 & 0 & 0 & 0 & 0 & -1 & -\sqrt{3} \\
 0 & 0 & 0 & 0 & 0 & 0 & \sqrt{3} & -1 \\
 0 & 0 & 0 & 0 & -1 & -\sqrt{3} & 0 & 0 \\
 0 & 0 & 0 & 0 & -\sqrt{3} & 1 & 0 & 0 \\
\end{array}
\right)
, \quad \\
  \rho_{\mathbf{8}}(P) ~=&~ \left(
\begin{array}{cccccccc}
 0 & 0 & 0 & 0 & 1 & 0 & 0 & 0 \\
 0 & 0 & 0 & 0 & 0 & 1 & 0 & 0 \\
 0 & 0 & 0 & 0 & 0 & 0 & 1 & 0 \\
 0 & 0 & 0 & 0 & 0 & 0 & 0 & -1 \\
 1 & 0 & 0 & 0 & 0 & 0 & 0 & 0 \\
 0 & -1 & 0 & 0 & 0 & 0 & 0 & 0 \\
 0 & 0 & 1 & 0 & 0 & 0 & 0 & 0 \\
 0 & 0 & 0 & 1 & 0 & 0 & 0 & 0 \\
\end{array}
\right).
\end{align}
This group is ambivalent, meaning that each element is in the same conjugacy class as its inverse.
Due to this fact, any involutory inner automorphism is class-inverting and, hence, can be used to define a proper physical CP transformation~\cite{Chen:2014tpa}. 
In fact, since outer automorphisms are cosets of the inner automorphisms all choices of inner automorphisms are identified, 
and the simplest choice for a physical CP transformation is just the trivial outer accompanied by the trivial inner automorphism.
The twisted Frobenius-Schur indicator (FSI) with respect to the trivial automorphism is displayed in Tab.~\ref{tab:FBI_Sig72}.
Since the twisted FSI with respect to the trivial automorphism corresponds to the regular FSI, 
this also shows the pseudo-reality of the doublet irrep.
\begin{table}
\centering
\begin{tabular}{ccccccc}
 $\mathbf{r}_j$ & $\mathbf{1}_0$ & $\mathbf{1}_1$ & $\mathbf{1}_2$ & $\mathbf{1}_3$ & $\mathbf{2}$ & $\mathbf{8}$ \\
\hline
$\text{FS}_{\mathrm{id}}(\mathbf{r}_j)$ & $1$ & $1$ & $1$ & $1$ & $-1$ & $1$
\end{tabular}
\caption{\label{tab:FBI_Sig72}%
Twisted Frobenius-Schur indicator (FSI) with respect to the trivial automorphism of $\Sigma(72)$.}
\end{table}
The thereby defined CP transformation acts on the irreps as shown in Eq.~\eqref{eq:CP72MapRule}. 
The fact that \rep{2} has an FSI of $-1$ is the very reason we identify 
$\Sigma(72)$ as a type~II~B group, and it also implies the necessarily antisymmetry $U_{\rep{2}}=-U^{\mathrm{T}}_{\rep{2}}$ of the doublet's 
CP transformation matrix. This also implies that there is \textit{no} basis in which all Clebsch-Gordan coefficients of 
$\Sigma(72)$ are real, see \cite{Chen:2014tpa,Trautner:2016ezn} for details. 

Finally, we explicitly state the generators of the regular representation in the permutation basis.
$M$ and $P$ are represented by permutation matrices that map the list of group elements $(1,2,3,\dots,72)$ (in the native ordering of GAP) 
to the lists
\begin{align}\label{eq:G72inElBasis}
    \pi_{\mathrm{reg}}(M) = & \left(3, 19, 4, 11, 12, 13, 2, 7, 37, 38, 1, 14, 15, 27, 28, 29, 30, 31, 8, 9, 10, 20,\right. \nonumber\\\nonumber
              &21, 55, 56, 57, 5, 6, 32, 33, 34, 47, 48, 49, 50, 51, 22, 23, 24, 25, 26, 39, 40,\\\nonumber
              &41, 67, 68, 16, 17, 18, 52, 53, 63, 64, 65, 42, 43, 44, 45, 46, 58, 59, 72, 35, 36, \\\nonumber
              &\left.66, 71, 60, 61, 62, 69, 54, 70\right)\;,\\\nonumber
    \pi_{\mathrm{reg}}(P) = &\left(9, 53, 21, 42, 24, 25, 48, 35, 66, 14, 57, 40, 41, 8, 60, 2, 45, 46, 65, 63, 64,\right. \\\nonumber
    &6, 54, 34, 32, 33, 68, 19, 58, 59, 7, 22, 23, 70, 10, 62, 31, 29, 28, 71, 27, 17, \\\nonumber
    &18, 16, 4, 52, 72, 37, 38, 69, 20, 43, 44, 26, 51, 3, 50, 49, 47, 36, 1, 15, 55, 56, \\
    &\left. 39, 61, 12, 13, 11, 5, 67, 30\right)\;.
\end{align}
Analogously, the generators of the conjugate representation in the permutation basis are given as
\begin{align}
    \pi_{\text{conj}}(M) = &\left(1, 8, 3, 4, 18, 5, 19, 2, 44, 22, 11, 31, 12, 34, 14, 6, 36, 16, 7, 57, 37, 26, 9,\right.\nonumber\\\nonumber
              &23, 61, 42, 49, 27, 13, 51, 29, 15, 53, 32, 17, 54, 41, 20, 38, 68, 55, 10, 46, 24, \\\nonumber
              &43, 70, 28, 64, 47, 30, 65, 33, 66, 35, 21, 59, 39, 56, 72, 25, 62, 60, 48, 71, 50, \\\nonumber
              &\left.52, 40, 69, 67, 45, 63, 58\right)\;, \\\nonumber
    \pi_{\text{conj}}(P) = &\left(1, 62, 47, 52, 35, 17, 56, 44, 9, 2, 13, 48, 28, 53, 34, 36, 18, 54, 20, 57, 72, 42,\right. \nonumber\\\nonumber 
              &22, 10, 45, 25, 65, 51, 49, 71, 64, 4, 32, 14, 16, 5, 21, 58, 55, 37, 19, 43, 23, \\\nonumber
              &60, 46, 26, 12, 3, 29, 27, 11, 33, 15, 6, 69, 59, 41, 38, 67, 70, 61, 24, 50, 30, \\
              &\left. 63, 66, 7, 39, 68, 8, 31, 40\right)\;.
\end{align}

\bibliography{bibliography}
\bibliographystyle{utphys}

\end{document}